\PassOptionsToPackage{unicode}{hyperref}
\PassOptionsToPackage{hyphens}{url}
\PassOptionsToPackage{dvipsnames,svgnames,x11names}{xcolor}
\documentclass[
  conference,
  twocolumn]{IEEEtran}
\usepackage{xcolor}
\usepackage{amsmath,amssymb}
\usepackage{iftex}
\ifPDFTeX
  \usepackage[T1]{fontenc}
  \usepackage[utf8]{inputenc}
  \usepackage{textcomp} % provide euro and other symbols
\else % if luatex or xetex
  \usepackage{unicode-math} % this also loads fontspec
  \defaultfontfeatures{Scale=MatchLowercase}
  \defaultfontfeatures[\rmfamily]{Ligatures=TeX,Scale=1}
\fi
\usepackage{lmodern}
\ifPDFTeX\else
\fi
\IfFileExists{upquote.sty}{\usepackage{upquote}}{}
\IfFileExists{microtype.sty}{% use microtype if available
  \usepackage[]{microtype}
  \UseMicrotypeSet[protrusion]{basicmath} % disable protrusion for tt fonts
}{}
\makeatletter
\@ifundefined{KOMAClassName}{% if non-KOMA class
  \IfFileExists{parskip.sty}{%
    \usepackage{parskip}
  }{% else
    \setlength{\parindent}{0pt}
    \setlength{\parskip}{6pt plus 2pt minus 1pt}}
}{% if KOMA class
  \KOMAoptions{parskip=half}}
\makeatother
\makeatletter
\ifx\paragraph\undefined\else
  \let\oldparagraph\paragraph
  \renewcommand{\paragraph}{
    \@ifstar
      \xxxParagraphStar
      \xxxParagraphNoStar
  }
  \newcommand{\xxxParagraphStar}[1]{\oldparagraph*{#1}\mbox{}}
  \newcommand{\xxxParagraphNoStar}[1]{\oldparagraph{#1}\mbox{}}
\fi
\ifx\subparagraph\undefined\else
  \let\oldsubparagraph\subparagraph
  \renewcommand{\subparagraph}{
    \@ifstar
      \xxxSubParagraphStar
      \xxxSubParagraphNoStar
  }
  \newcommand{\xxxSubParagraphStar}[1]{\oldsubparagraph*{#1}\mbox{}}
  \newcommand{\xxxSubParagraphNoStar}[1]{\oldsubparagraph{#1}\mbox{}}
\fi
\makeatother

\usepackage{longtable,booktabs,array}
\usepackage{calc} % for calculating minipage widths
\usepackage{etoolbox}
\makeatletter
\patchcmd\longtable{\par}{\if@noskipsec\mbox{}\fi\par}{}{}
\makeatother
\IfFileExists{footnotehyper.sty}{\usepackage{footnotehyper}}{\usepackage{footnote}}
\makesavenoteenv{longtable}
\usepackage{graphicx}
\makeatletter
\newsavebox\pandoc@box
\newcommand*\pandocbounded[1]{% scales image to fit in text height/width
  \sbox\pandoc@box{#1}%
  \Gscale@div\@tempa{\textheight}{\dimexpr\ht\pandoc@box+\dp\pandoc@box\relax}%
  \Gscale@div\@tempb{\linewidth}{\wd\pandoc@box}%
  \ifdim\@tempb\p@<\@tempa\p@\let\@tempa\@tempb\fi% select the smaller of both
  \ifdim\@tempa\p@<\p@\scalebox{\@tempa}{\usebox\pandoc@box}%
  \else\usebox{\pandoc@box}%
  \fi%
}
\def\fps@figure{htbp}
\makeatother

\NewDocumentCommand\citeproctext{}{}

\makeatletter
 \let\@cite@ofmt\@firstofone
 \def\@biblabel#1{}
 \def\@cite#1#2{{#1\if@tempswa , #2\fi}}
\makeatother
\newlength{\cslhangindent}
\newlength{\csllabelwidth}
\newenvironment{CSLReferences}[2] % #1 hanging-indent, #2 entry-spacing
 {\begin{list}{}{%
  \setlength{\itemindent}{0pt}
  \setlength{\leftmargin}{0pt}
  \setlength{\parsep}{0pt}
  \ifodd #1
   \setlength{\leftmargin}{\cslhangindent}
   \setlength{\itemindent}{-1\cslhangindent}
  \fi
  \setlength{\itemsep}{#2\baselineskip}}}
 {\end{list}}
\usepackage{calc}

\providecommand{\tightlist}{%
  \setlength{\itemsep}{0pt}\setlength{\parskip}{0pt}}

\makeatletter
\@ifpackageloaded{caption}{}{\usepackage{caption}}
\AtBeginDocument{%
\ifdefined\contentsname
  \renewcommand*\contentsname{Table of contents}
\else
  \newcommand\contentsname{Table of contents}
\fi
\ifdefined\listfigurename
  \renewcommand*\listfigurename{List of Figures}
\else
  \newcommand\listfigurename{List of Figures}
\fi
\ifdefined\listtablename
  \renewcommand*\listtablename{List of Tables}
\else
  \newcommand\listtablename{List of Tables}
\fi
\ifdefined\figurename
  \renewcommand*\figurename{Figure}
\else
  \newcommand\figurename{Figure}
\fi
\ifdefined\tablename
  \renewcommand*\tablename{Table}
\else
  \newcommand\tablename{Table}
\fi
}
\@ifpackageloaded{float}{}{\usepackage{float}}
\floatstyle{ruled}
\@ifundefined{c@chapter}{\newfloat{codelisting}{h}{lop}}{\newfloat{codelisting}{h}{lop}[chapter]}
\floatname{codelisting}{Listing}

\makeatother
\makeatletter
\@ifpackageloaded{caption}{}{\usepackage{caption}}
\@ifpackageloaded{subcaption}{}{\usepackage{subcaption}}
\makeatother
\usepackage{bookmark}
\IfFileExists{xurl.sty}{\usepackage{xurl}}{} % add URL line breaks if available
\hypersetup{
  pdftitle={Physics-Informed Diffusion Models for Vehicle Speed Trajectory Generation},
  pdfauthor={Vadim Sokolov; Farnaz Behnia; Dominik Karbowski},
  pdfkeywords={Data-based approaches (learning, deep learning,
reinforcement learning), Energy management methods, Road
transportation, Connected and Autonomous Vehicles, speed
generation, diffusion models},
  colorlinks=true,
  linkcolor={blue},
  filecolor={Maroon},
  citecolor={Blue},
  urlcolor={Blue},
  pdfcreator={LaTeX via pandoc}}

\title{Physics-Informed Diffusion Models for Vehicle Speed Trajectory
Generation}
\author{
\IEEEauthorblockN{Vadim Sokolov}\\
\IEEEauthorblockA{George Mason University}
\and
\IEEEauthorblockN{Farnaz Behnia}\\
\IEEEauthorblockA{George Mason University}
\and
\IEEEauthorblockN{Dominik Karbowski}\\
\IEEEauthorblockA{Argonne National Laboratory}
}
\date{}
\begin{document}
\maketitle
\begin{abstract}
Synthetic vehicle speed trajectory generation is essential for
evaluating vehicle control algorithms and connected vehicle
technologies. Traditional Markov chain approaches suffer from
discretization artifacts and limited expressiveness. This paper proposes
a physics-informed diffusion framework for conditional micro-trip
synthesis, combining a dual-channel speed-acceleration representation
with soft physics constraints that resolve optimization conflicts
inherent to hard-constraint formulations. We compare a 1D U-Net
architecture against a transformer-based Conditional Score-based
Diffusion Imputation (CSDI) model using 6,367 GPS-derived micro-trips.
CSDI achieves superior distribution matching (Wasserstein distance 0.30
for speed, 0.026 for acceleration), strong indistinguishability from
real data (discriminative score 0.49), and validated utility for
downstream energy assessment tasks. The methodology enables scalable
generation of realistic driving profiles for intelligent transportation
systems (ITS) applications without costly field data collection.
\end{abstract}

\section{Introduction}\label{sec-introduction}

Accurate modeling of vehicle speed trajectories is fundamental to
numerous applications in intelligent transportation systems (ITS),
including energy consumption assessment (Karbowski, Sokolov, and
Rousseau 2015; Karbowski et al. 2016; Karbowski, Sokolov, and Jongryeol
2016), route optimization (Moawad et al. 2021), traffic simulation (Auld
et al. 2013; Auld, Hope, et al. 2016), and predictive control of
autonomous vehicles (Dauner et al. 2023). The ability to generate
synthetic speed profiles that faithfully reproduce the statistical and
physical characteristics of real-world driving behavior enables scalable
evaluation of transportation policies, vehicle technologies, and traffic
management strategies without the prohibitive cost of extensive field
data collection.

Traditional approaches to synthetic trajectory generation have relied on
Markov chain models (Karbowski et al., n.d., 2016), which discretize the
speed state space and model transitions between consecutive states.
While computationally efficient and interpretable, these methods face
fundamental limitations: rigid discretization loses fine-grained
dynamics, the Markov assumption ignores long-range temporal
dependencies, and incorporating physics-based constraints or conditional
controls (e.g., powertrain type, road conditions) requires ad-hoc
engineering. More recent efforts have explored deep generative models,
including generative adversarial networks (GANs) (Behnia, Karbowski, and
Sokolov 2023; Yoon, Jarrett, and van der Schaar 2019) and normalizing
flows (Papamakarios et al. 2019), with mixed success due to training
instability, mode collapse, and difficulty enforcing hard boundary
constraints.

This paper introduces an application of denoising diffusion
probabilistic models (DDPMs) (Ho, Jain, and Abbeel 2020) to the problem
of conditional vehicle speed trajectory synthesis. Diffusion models have
emerged as an effective class of generative models, demonstrating strong
performance in image generation, audio synthesis, and recently, time
series forecasting (Ansari et al. 2025). Their key advantages include
stable training dynamics, the ability to naturally model complex
multi-modal distributions, and relevant for this application, a
principled mechanism for enforcing constraints through inpainting during
the reverse diffusion process. Unlike prior work on diffusion-based
trajectory generation for autonomous driving (Feng et al. 2023; Suo et
al. 2021), which focuses on spatial path planning in multi-agent
scenarios, we target univariate speed profiles for energy assessment
applications, requiring strict enforcement of micro-trip boundaries
(zero initial and final speeds) and precise control over aggregate
statistics (average speed, duration).

We compare two diffusion-based architectures: a standard 1D U-Net
diffusion model and a Conditional Score-based Diffusion Imputation
(CSDI) model (Tashiro et al. 2021) adapted from time series imputation.
Both models are conditioned on trip characteristics (target average
speed, duration) and trained on 6,367 micro-trip observations from the
2007 Chicago Metropolitan Agency for Planning (CMAP) Regional Household
Travel Survey. Our investigation systematically explores the evolution
of these models through multiple design iterations, documenting the
failures of hard-constraint physics penalties in standard diffusion and
the successful integration of soft physics constraints in CSDI. We
benchmark these approaches against traditional Markov chain baselines
and analyze the failure modes of alternative deep generative methods
(DoppelGANger, SDV) to provide guidance for practitioners.

The practical utility of high-fidelity synthetic trajectory generation
spans several key areas of intelligent transportation systems. For
energy assessment, synthetic trajectories enable evaluation of electric
vehicle range, charging infrastructure requirements, and fleet
electrification strategies across diverse driving scenarios without
costly field trials (Chen et al. 2016; X. Huang et al. 2020). Traffic
microsimulation platforms like POLARIS (Auld, Hope, et al. 2016) require
realistic speed profiles to accurately model emissions, fuel
consumption, and network-level energy impacts. Connected and autonomous
vehicle testing demands diverse, physically plausible trajectories for
validating perception, planning, and control systems (Mozaffari et al.
2020), as well as generating representative safety-critical scenarios
(Wu et al. 2025). By demonstrating both distributional fidelity (via
Wasserstein metrics) and downstream utility (via TSTR), this work
provides a validated methodology for generating synthetic driving data
that directly supports ITS energy management, simulation-based policy
analysis, and intelligent vehicle development.

The principal contributions of this work are:

\begin{enumerate}
\def\labelenumi{\arabic{enumi}.}
\tightlist
\item
  The first application of diffusion models specifically to vehicle
  speed micro-trip generation for transportation energy assessment,
  demonstrating superior distribution matching compared to Markov chains
  and GANs.
\item
  The first successful integration of kinematic physics constraints into
  transformer-based diffusion for transportation, combining direct speed
  and acceleration channels with physics-based training objectives to
  effectively bridge the gap between deep generative modeling and
  kinematic consistency.
\item
  Documentation of the complete model development lifecycle, including
  failures (hard-constraint diffusion, DoppelGANger, SDV), providing
  insights into why certain approaches fail and how to avoid common
  pitfalls in generative modeling for transportation applications.
\item
  A rigorous evaluation framework encompassing distributional fidelity
  (Wasserstein distance, MMD), kinematic validity (smoothness metrics,
  boundary violations), and utility (discriminative score, TSTR),
  establishing best practices for assessing synthetic trajectory
  quality.
\item
  Open-source implementation and reproducibility package at
  \url{https://github.com/VadimSokolov/diffusion-trajectory-generation},
  including complete training code, pretrained model weights,
  preprocessed data, and figure generation scripts enabling full
  reproduction of all results.
\end{enumerate}

The remainder of this paper is organized as follows:
Section~\ref{sec-related-work} reviews related work in vehicle
trajectory generation, deep generative models for time series, and
diffusion models. Section~\ref{sec-data} describes the CMAP dataset,
preprocessing, and clustering analysis. Section~\ref{sec-methodology}
presents the methodology, including problem formulation, the Markov
chain baseline, and detailed descriptions of the diffusion and CSDI
architectures along with their evolution through multiple design
iterations. Section~\ref{sec-results} reports quantitative and
qualitative results, comparing all approaches and documenting failure
modes. Section~\ref{sec-discussion} discusses implications for practice,
limitations, and future research directions.

\section{Related Work}\label{sec-related-work}

The problem of generating realistic vehicle trajectories has been
approached from multiple perspectives in the transportation literature.
Early work relied on parametric models and Markov processes to capture
driving patterns (Karbowski et al., n.d., 2016), where speed transitions
are modeled as discrete-state stochastic processes. Karbowski et
al.~developed Markov chain models for trip prediction in
energy-efficient vehicle routing, demonstrating that second-order Markov
models can capture speed-acceleration dependencies (Karbowski, Sokolov,
and Jongryeol 2016). While interpretable and computationally efficient,
these approaches are limited by their memoryless nature and inability to
model long-range temporal dependencies inherent in driving behavior.

Agent-based traffic simulation platforms such as POLARIS (Sokolov, Auld,
and Hope 2012; Auld et al. 2013; Auld, Hope, et al. 2016) have
incorporated microsimulation of individual vehicle movements to assess
regional transportation policies. These systems combine activity-based
demand models with network assignment, requiring realistic speed
profiles as inputs (Auld, Karbowski, et al. 2016). Moawad et
al.~proposed neural recommender systems for real-time route assignment
considering energy consumption, highlighting the need for fast
generation of representative speed trajectories conditioned on route
characteristics (Moawad et al. 2021).

More recently, Behnia et al.~explored deep generative models for vehicle
speed trajectories, comparing variational autoencoders (VAEs), GANs, and
normalizing flows (Behnia, Karbowski, and Sokolov 2023). Their analysis
demonstrated that GANs achieved the best distribution matching but
suffered from training instability and mode collapse, particularly for
diverse driving regimes. This work established important evaluation
metrics for trajectory generation but did not explore diffusion models,
which have since emerged as a more stable alternative to GANs.

Adjacent to single-vehicle speed generation is the rapidly growing
literature on multi-agent trajectory prediction and scenario generation
for autonomous vehicle testing. TrafficGen (Feng et al. 2023) and
ScenarioNet (Li et al. 2023) introduced learning-based frameworks for
generating diverse, realistic traffic scenarios from large-scale driving
datasets. TrafficSim (Suo et al. 2021) proposed a multi-agent behavioral
model learning to simulate realistic interactions among vehicles.

These works differ fundamentally from our focus: they generate spatial
trajectories (x, y coordinates over time) for multiple interacting
agents in 2D road networks, whereas we target univariate speed profiles
for isolated micro-trips. The former requires modeling complex spatial
interactions and collision avoidance; the latter emphasizes matching
kinematic distributions and enforcing strict boundary conditions for
energy assessment. Nevertheless, recent applications of diffusion models
to autonomous driving trajectory planning (Liao et al. 2024) and
controllable motion generation (Lan et al. 2025) demonstrate the
potential of this generative modeling paradigm for transportation
applications.

The application of deep generative models to sequential data has evolved
rapidly. TimeGAN (Yoon, Jarrett, and van der Schaar 2019) extended GANs
to time series by combining unsupervised adversarial learning with
supervised stepwise prediction, improving temporal coherence.
DoppelGANger addressed the challenge of generating mixed-type time
series with both temporal features and static attributes through a
dual-stage generation process. However, both methods remain susceptible
to mode collapse and require careful architectural tuning to enforce
constraints.

Autoregressive models, including probabilistic variants implemented in
the Synthetic Data Vault (SDV), represent another popular approach. The
PARSynthesizer uses probabilistic autoregression to sequentially
generate time steps conditioned on previous values. While simple to
implement, these models suffer from exposure bias---errors accumulate
over long sequences---and struggle with capturing diverse multi-modal
distributions without explicit diversity mechanisms.

Foundation models for time series have recently emerged as zero-shot
alternatives. Chronos (Ansari et al. 2025) adapts transformer language
model architectures to probabilistic forecasting by tokenizing time
series and training on diverse datasets. While showing promise for
forecasting tasks, these models lack fine-grained control over
generation and have not been extensively evaluated for constrained
trajectory synthesis.

Denoising diffusion probabilistic models (DDPMs) (Ho, Jain, and Abbeel
2020) define a forward process that gradually corrupts data with
Gaussian noise and learn a reverse process to denoise random samples
into data. Unlike GANs, diffusion models optimize a tractable
variational lower bound, resulting in stable training and high sample
quality. Score-based generative models (Y. Song et al. 2020) provide an
equivalent formulation by learning the gradient of the data log-density
(score function) and using Langevin dynamics for sampling.

The application of diffusion models to time series has largely focused
on forecasting and imputation. CSDI (Conditional Score-based Diffusion
for Imputation) (Tashiro et al. 2021) introduced a transformer-based
diffusion architecture for missing value imputation in multivariate time
series, demonstrating superior performance to autoregressive and
GAN-based methods. The key innovation is self-attention's ability to
capture long-range dependencies while the diffusion framework handles
uncertainty. We adapt this architecture for conditional generation by
treating the entire trajectory as ``missing'' conditioned on aggregate
trip statistics. Recent extensions in the ITS domain have combined
diffusion with GANs for trajectory reconstruction (Qian et al. 2025) and
proposed interpretable causal diffusion networks for speed prediction
(Rong et al. 2025).

TimeGrad (Rasul et al. 2021) introduced autoregressive denoising
diffusion for probabilistic time series forecasting, demonstrating
strong performance on multivariate prediction tasks. SSSD (Alcaraz and
Strodthoff 2022) combined structured state-space models with diffusion
for imputation and forecasting. TimeWeaver and Diffusion-TS have further
extended diffusion models to multivariate time series synthesis.
However, these methods do not address the specific challenges of vehicle
trajectory generation: strict boundary constraints (zero start/end
speeds), physics-based plausibility (acceleration limits, smoothness),
and fine-grained conditional control (vehicle type, road conditions).
While deep learning has been extensively applied to trajectory
prediction in ITS (Schultz and Sokolov 2018; Mozaffari et al. 2020; Zhao
et al. 2019; Altché and de La Fortelle 2017; Y. Huang et al. 2022),
generation of complete speed profiles for energy assessment remains
underexplored. Multi-agent diffusion models have also been explored for
traffic flow prediction (Adam et al. 2025).

To our knowledge, this work is the first to apply diffusion models
specifically to vehicle speed micro-trip generation for energy
assessment, systematically comparing U-Net and transformer-based
architectures and documenting the integration of physics-informed
constraints.

\section{Data}\label{sec-data}

The data for this study originate from the 2007 Chicago Metropolitan
Agency for Planning (CMAP) Regional Household Travel Survey, conducted
by NuStats for CMAP, an 11-county region, encompassing dense urban
(Chicago), suburban, and rural environments. This geographic diversity
ensures representation of varied driving conditions, road types, and
traffic patterns.

The survey employed a dual-frame sampling strategy combining Random
Digit Dialing (RDD) and address-based sampling to reach over 3.2 million
households. Data collection utilized Computer-Assisted Telephone
Interview (CATI), mail questionnaires, and follow-up telephone
interviews, achieving an overall response rate of 10 percent (19 percent
recruitment rate, 55 percent retrieval rate) (Chicago Metropolitan
Agency for Planning 2008). A subset of participating households
contributed GPS-based trip-level data, which forms the basis of our
analysis.

The dataset captures passenger vehicle movements (cars, SUVs, vans,
pickup trucks) used for personal travel. Heavy-duty commercial vehicles
and fleet operations are excluded. Road type coverage includes
Interstate highways and expressways, major arterials connecting economic
centers, and local urban and suburban roads. No filtering by road type
was applied during collection; GPS traces reflect the natural
distribution of household travel patterns.

\subsection{Exploratory Data Analysis}\label{exploratory-data-analysis}

Following preprocessing to extract individual micro-trips (continuous
speed profiles beginning and ending at zero velocity), our dataset
comprises 6,367 observations sampled at 1 Hz.
Table~\ref{tbl-dataset-stats} summarizes key distributional statistics.

\begin{table}[htbp]
\centering
\caption{Dataset Summary Statistics}\label{tbl-dataset-stats}
\begin{tabular}{@{}
lrrrr@{}}

Attribute & Min & Max & Mean & Median \\
\midrule\noalign{}
Duration (s) & 34 & 12,841 & 304 & 187 \\
Duration (min) & 0.57 & 214 & 5.06 & 3.12 \\
Distance (m) & 257 & 393,070 & 5,884 & 3,076 \\
Distance (km) & 0.26 & 393 & 5.88 & 3.08 \\
Avg Speed (m/s) & 5.18 & 31.57 & 16.92 & 16.45 \\
Avg Speed (km/h) & 18.63 & 113.66 & 60.92 & 59.21 \\
\end{tabular}
\end{table}

The dataset exhibits substantial variation in trip characteristics, with
durations ranging from half a minute to over three hours and distances
spanning two orders of magnitude. Average speeds vary from slow urban
crawl (5 m/s \(\approx\) 18 km/h) to highway cruising (32 m/s
\(\approx\) 115 km/h), reflecting the diversity of driving contexts in
the metropolitan area.

\begin{figure}

\centering{

\includegraphics[width=1\linewidth,height=\textheight,keepaspectratio]{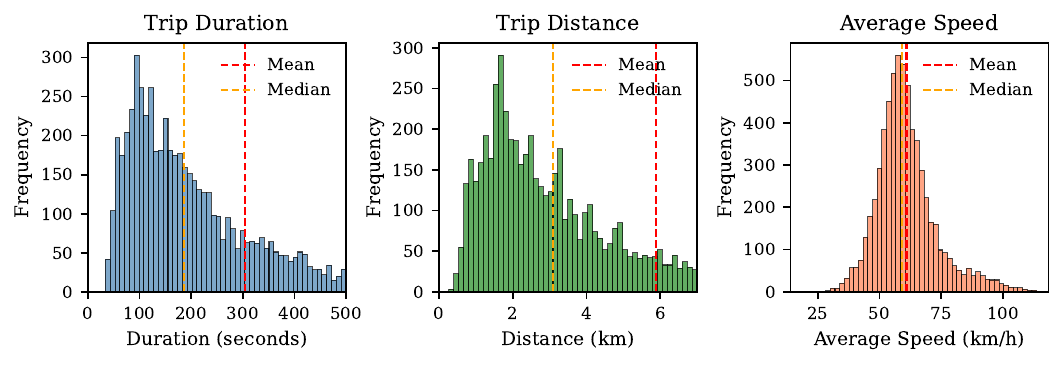}

}

\caption{\label{fig-data-dist}Data distributions for trip duration,
distance, and average speed. Histograms show the empirical distributions
from 6,367 micro-trips with mean (red dashed) and median (orange dashed)
markers. The distributions exhibit substantial heterogeneity, reflecting
diverse driving contexts from urban congestion to highway cruising.}

\end{figure}%

To characterize the heterogeneity of driving patterns in the dataset, we
performed K-means clustering (K=4) on extracted features from each
trajectory. Clustering serves two purposes: (1) understanding the data
structure to inform model design, and (2) enabling stratified evaluation
to ensure generated trajectories cover all driving regimes.

\emph{Feature Extraction}: For each micro-trip, we computed:

\begin{itemize}
\tightlist
\item
  Average speed (\(\bar{v}\)) and maximum speed (\(v_{max}\))
\item
  Speed standard deviation
\item
  Idle time ratio (fraction of time at \(v < 0.5\) m/s)
\item
  Stops per kilometer (number of near-zero speed events per unit
  distance)
\item
  Acceleration noise (standard deviation of acceleration)
\end{itemize}

\emph{Clustering Results}: The algorithm identified four distinct
driving regimes, described in Table~\ref{tbl-cluster-chars}:

\begin{table}[htbp]
\centering
\caption{Cluster Characteristics}\label{tbl-cluster-chars}
\begin{tabular}{@{}
>{\raggedright\arraybackslash}p{(\linewidth - 12\tabcolsep) * \real{0.1000}}
  >{\raggedright\arraybackslash}p{(\linewidth - 12\tabcolsep) * \real{0.2500}}
  >{\raggedleft\arraybackslash}p{(\linewidth - 12\tabcolsep) * \real{0.0700}}
  >{\raggedleft\arraybackslash}p{(\linewidth - 12\tabcolsep) * \real{0.1600}}
  >{\raggedleft\arraybackslash}p{(\linewidth - 12\tabcolsep) * \real{0.1600}}
  >{\raggedleft\arraybackslash}p{(\linewidth - 12\tabcolsep) * \real{0.1300}}
  >{\raggedleft\arraybackslash}p{(\linewidth - 12\tabcolsep) * \real{0.1300}}@{}}

\begin{minipage}[b]{\linewidth}\raggedright
Cluster
\end{minipage} & \begin{minipage}[b]{\linewidth}\raggedright
Label
\end{minipage} & \begin{minipage}[b]{\linewidth}\raggedleft
Count
\end{minipage} & \begin{minipage}[b]{\linewidth}\raggedleft
Avg Speed (m/s)
\end{minipage} & \begin{minipage}[b]{\linewidth}\raggedleft
Max Speed (m/s)
\end{minipage} & \begin{minipage}[b]{\linewidth}\raggedleft
Stops/km
\end{minipage} & \begin{minipage}[b]{\linewidth}\raggedleft
Idle Ratio (\%)
\end{minipage} \\
\midrule\noalign{}
0 & Arterial/ Suburban & 2,224 & 15.6 & 22.6 & 0.59 & 2.0 \\
1 & Highway/ Interstate & 1,020 & 22.2 & 30.8 & 0.12 & 0.5 \\
2 & Congested/ City & 636 & 13.9 & 21.8 & 1.29 & 4.4 \\
3 & Free-flow Arterial & 2,487 & 16.7 & 22.7 & 0.28 & 1.0 \\
\end{tabular}
\end{table}

Cluster 1 (Highway/Interstate) exhibits the highest average and maximum
speeds with minimal stops and idle time, characteristic of uninterrupted
highway travel. Cluster 2 (Congested/City) shows the opposite profile:
low speeds, frequent stops (1.29 per km), and high idle time (4.4
percent), typical of stop-and-go urban driving. Clusters 0 and 3
represent intermediate arterial road conditions, with Cluster 3 having
somewhat higher speeds and fewer stops, suggesting less congested
arterial travel.

\begin{figure}

\begin{minipage}{0.50\linewidth}

\centering{

\pandocbounded{\includegraphics[keepaspectratio]{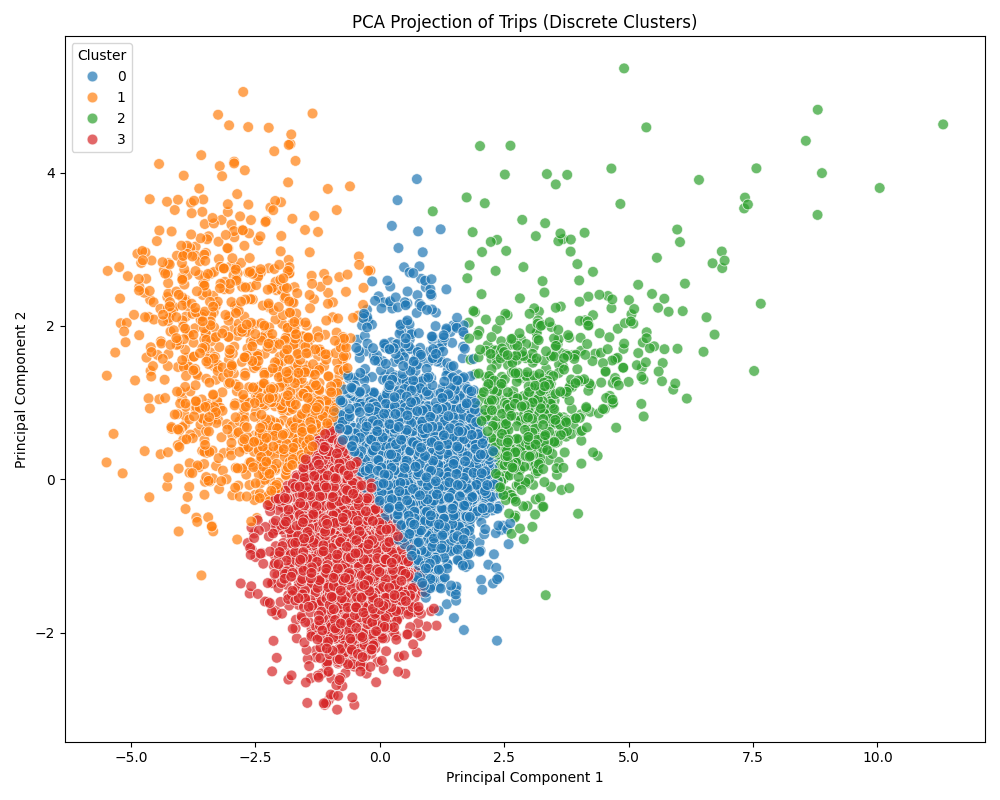}}

}

\subcaption{\label{fig-pca}PCA}

\end{minipage}%
\begin{minipage}{0.50\linewidth}

\centering{

\pandocbounded{\includegraphics[keepaspectratio]{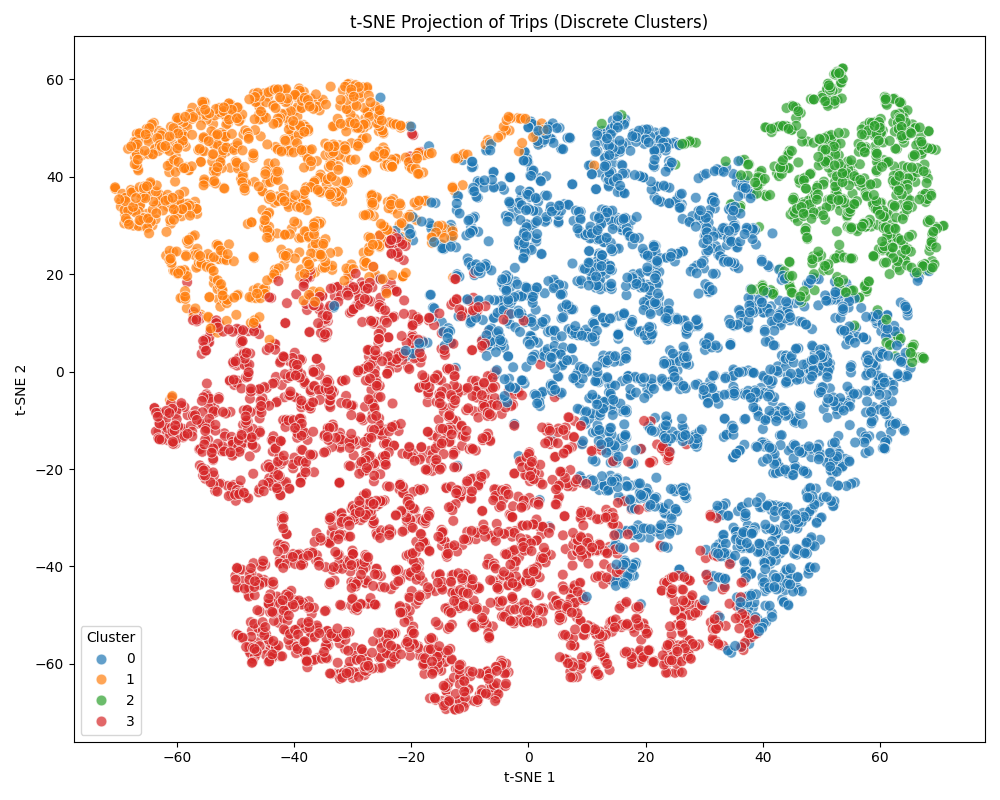}}

}

\subcaption{\label{fig-tsne}t-SNE}

\end{minipage}%

\caption{\label{fig-cluster-projections}PCA and t-SNE projections of
micro-trips in feature space.}

\end{figure}%

\begin{figure}

\begin{minipage}{0.50\linewidth}

\centering{

\pandocbounded{\includegraphics[keepaspectratio]{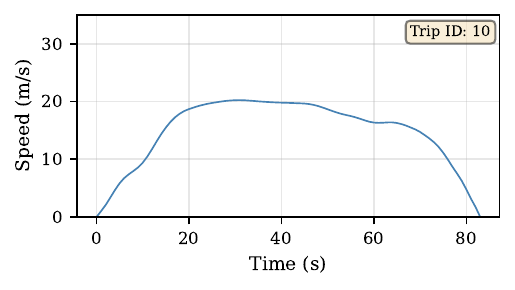}}

}

\subcaption{\label{fig-cluster0}Arterial, Cluster 0.}

\end{minipage}%
\begin{minipage}{0.50\linewidth}

\centering{

\pandocbounded{\includegraphics[keepaspectratio]{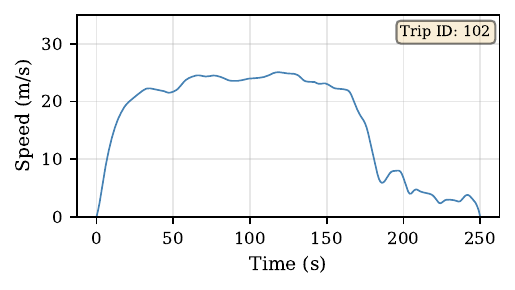}}

}

\subcaption{\label{fig-cluster1}Highway, Cluster 1.}

\end{minipage}%
\newline
\begin{minipage}{0.50\linewidth}

\centering{

\pandocbounded{\includegraphics[keepaspectratio]{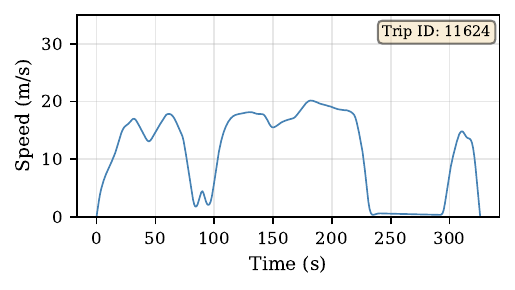}}

}

\subcaption{\label{fig-cluster2}City congested, Cluster 2.}

\end{minipage}%
\begin{minipage}{0.50\linewidth}

\centering{

\pandocbounded{\includegraphics[keepaspectratio]{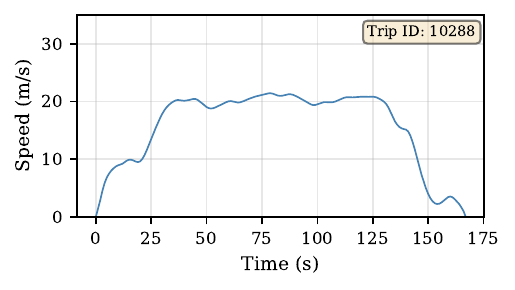}}

}

\subcaption{\label{fig-cluster3}City free-flow, Cluster 3.}

\end{minipage}%

\caption{\label{fig-cluster}Examples of observed trajectories from each
cluster.}

\end{figure}

Principal component analysis (PCA) and t-SNE projections
(Figure~\ref{fig-cluster-projections}) visualize the cluster separation
in feature space, with the first two PCA components explaining
approximately 65\% of the variance. t-SNE reveals a tighter cluster
structure that highlights the distinct kinematic signatures of the four
driving regimes. While clusters form distinguishable groups, the
continuous transitions between them reflect the spectrum of real-world
driving conditions rather than discrete categories. Relatedly,
Figure~\ref{fig-cluster} display example speed profiles from each
cluster, selected from the cluster-to-trip mapping to illustrate
characteristic patterns.

These clusters inform our evaluation strategy: we assess whether
generated trajectories match the proportions and characteristics of each
driving regime, ensuring the model captures both highway efficiency and
urban congestion patterns.

\section{Methodology}\label{sec-methodology}

Let \(\mathbf{v} = (v_0, v_1, \ldots, v_T) \in \mathbb{R}^{T+1}\)
represent a vehicle speed trajectory sampled at 1 Hz, where
\(t \in \{0, 1, \ldots, T\}\) indexes time in seconds. Our unit of
analysis is the \emph{micro-trip}: a stop-to-stop segment that begins
and ends at zero velocity. Each complete trip in the dataset is
partitioned into a sequence of such micro-trips, segmenting when the
vehicle comes to rest. This decomposition allows us to model the
fundamental building blocks of driving behavior; a complete
origin-to-destination journey can be reconstructed by concatenating
generated micro-trips. Each micro-trip satisfies the following
constraints:

\emph{Boundary conditions}:
\[v_0 = v_T = 0 \quad \text{(trip begins and ends at rest)}\]

\emph{Non-negativity}:
\[v_t \geq 0 \quad \forall t \in \{0, \ldots, T\}\]

\emph{Conditioning variables}: We denote the conditioning information as
\(\mathbf{c} \in \mathbb{R}^d\), typically comprising:

\begin{itemize}
\tightlist
\item
  \(\bar{v}\): target average speed (m/s)
\item
  \(T\): trip duration (seconds)
\item
  \(v_{max}\): maximum speed (m/s) {[}for CSDI only{]}
\item
  \(d_{veh}\): vehicle dynamics parameter \(\in [0,1]\) {[}for CSDI
  only{]}
\end{itemize}

The \emph{objective} is to learn a conditional generative model
\(p_\theta(\mathbf{v}|\mathbf{c})\) that:

\begin{enumerate}
\def\labelenumi{\arabic{enumi}.}
\tightlist
\item
  Generates trajectories matching the empirical distribution
  \(p_{data}(\mathbf{v})\)
\item
  Satisfies physical plausibility constraints (smooth acceleration,
  bounded jerk)
\item
  Precisely controls aggregate trip statistics via conditioning
  \(\mathbf{c}\)
\end{enumerate}

Formally, we seek to minimize the distributional divergence:
\[\min_\theta D_{KL}(p_{data}(\mathbf{v}|\mathbf{c}) \| p_\theta(\mathbf{v}|\mathbf{c}))\]
while enforcing boundary and kinematic constraints.

\subsection{Baseline: Markov Chain
Approach}\label{baseline-markov-chain-approach}

Traditional trajectory generation relies on discrete-state Markov models
(Karbowski et al., n.d.). We implement a second-order Markov chain
baseline following Karbowski et al.'s methodology as a reference point
for evaluating deep generative models.

\emph{State Space Discretization}: Speed is discretized into bins of
width \(\Delta v = 0.5\) m/s, creating bins
\(\{B_0, B_1, \ldots, B_K\}\) where
\(B_i = [i\Delta v, (i+1)\Delta v)\). Let \(s_t \in \{0,1,\ldots,K\}\)
denote the bin index at time \(t\).

\emph{Second-Order Markov Model}: The model assumes speed at time \(t\)
depends on the previous two time steps:
\[P(s_t | s_{t-1}, s_{t-2}) = \frac{C(s_{t-2}, s_{t-1}, s_t)}{\sum_{s'} C(s_{t-2}, s_{t-1}, s')}\]
where \(C(\cdot)\) denotes empirical counts from the training data. This
can be reformulated as a first-order Markov chain on pair-states
\(X_t = (s_{t-1}, s_t)\) with state space size \(K^2\).

\emph{Boundary-Constrained Sampling (Markov Bridge)}: To enforce
\(v_0 = v_T = 0\), we employ forward-backward sampling (Durham and
Gallant 2002). Backward messages \(\beta_t(x)\) represent the
probability of reaching the terminal state (bin 0) from state \(x\) at
time \(t\):
\[\beta_T(x) = \mathbb{1}[x_{\text{end}} = 0], \quad \beta_t(x) = \sum_{x'} P(x' | x) \beta_{t+1}(x')\]

Forward sampling at each step uses the modified transition
probabilities:
\[P_{\text{bridge}}(x_{t+1} | x_t) \propto P(x_{t+1} | x_t) \cdot \beta_{t+1}(x_{t+1})\]

\emph{Post-Processing}: To reduce discretization artifacts, we apply
Gaussian smoothing to the acceleration signal \(a_t = v_{t+1} - v_t\)
with a 5-point moving average, followed by integration to recover smooth
speed profiles.

\emph{Limitations}: The Markov baseline suffers from three fundamental
weaknesses: (1) rigid bin discretization loses fine-grained dynamics,
(2) limited temporal memory cannot capture long-range trip structure
(e.g., highway segments followed by urban navigation), and (3)
conditional control requires separate models per condition or rejection
sampling, both of which are inefficient.

All models were trained on an NVIDIA A100 GPU (40GB) using PyTorch 2.0.
Training utilized mixed-precision (FP16) computation to accelerate
convergence and reduce memory footprint. The Markov chain baseline
requires no GPU training, completing parameter estimation (transition
matrix construction) in under 5 minutes on a single CPU core.

\subsection{Conditional Diffusion Model with 1D
U-Net}\label{conditional-diffusion-model-with-1d-u-net}

Our diffusion model employs a 1D U-Net encoder-decoder architecture with
Feature-wise Linear Modulation (FiLM) for conditioning. The input is a
two-channel tensor \(\mathbf{x}_t \in \mathbb{R}^{2 \times 512}\)
representing the joint speed-acceleration state:
\[\mathbf{x}_0 = \begin{bmatrix} v_0, v_1, \ldots, v_{511} \\ a_0, a_1, \ldots, a_{511} \end{bmatrix}\]
where \(a_t = v_{t+1} - v_t\) is the discrete acceleration. Trajectories
shorter than 512 seconds are zero-padded; longer trajectories are
truncated (less than 1\% of data).

The \emph{encoder} consists of four downsampling blocks with channel
dimensions {[}64, 128, 256, 512{]}, each containing:

\begin{itemize}
\tightlist
\item
  1D convolution (kernel size 3, stride 2)
\item
  Two ResNet blocks with Group Normalization and SiLU activation
\item
  FiLM conditioning layers injecting time embedding and trip conditions
\end{itemize}

\emph{Self-attention} layers are inserted at resolutions 128 and 64 to
capture long-range dependencies without computational explosion at the
full 512 resolution.

The \emph{decoder} mirrors the encoder with skip connections, upsampling
to reconstruct the predicted noise
\(\boldsymbol{\epsilon}_\theta(\mathbf{x}_t, t, \mathbf{c}) \in \mathbb{R}^{2 \times 512}\).

\emph{Time Embedding}: The diffusion timestep
\(t \in \{1, \ldots, T_{diff}\}\) is embedded using sinusoidal
positional encoding:
\[\text{PE}_i(t) = \begin{cases} \sin(t / 10000^{i/d}) & i \text{ even} \\ \cos(t / 10000^{i/d}) & i \text{ odd} \end{cases}\]
projected to 256 dimensions.

\emph{Conditioning via FiLM}: At each ResNet block, the conditioning
vector \(\mathbf{c} = [\bar{v}/30,  T/1000]\) (normalized) is
transformed into scale and shift parameters using Feature-wise Linear
Modulation (FiLM) (Perez et al. 2018):
\[\text{FiLM}(\mathbf{h}, \mathbf{c}) = \gamma(\mathbf{c}) \odot \mathbf{h} + \beta(\mathbf{c})\]
where \(\gamma, \beta : \mathbb{R}^2 \to \mathbb{R}^{d_h}\) are learned
MLP projections. This multiplicative and additive modulation allows the
model to adapt its feature representations based on trip conditions.

\begin{figure}

\centering{

\includegraphics[width=3.5in,height=0.78in]{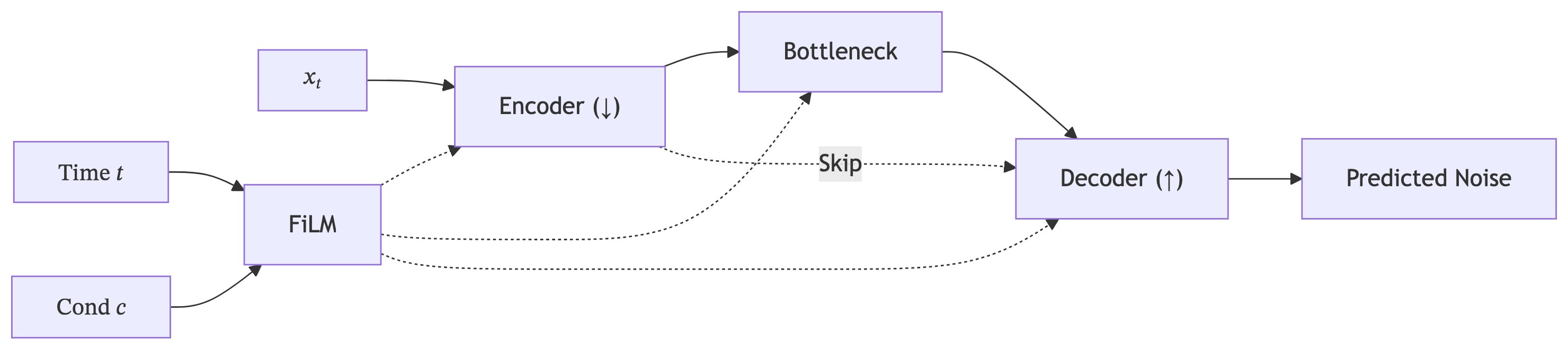}

}

\caption{\label{fig-unet-arch}Simplified U-Net Diffusion Architecture.
Noisy trajectory \(x_t\) is encoded to a latent representation and
decoded to predict noise, modulated by time \(t\) and conditions \(c\).}

\end{figure}%

\subsubsection{Forward Diffusion
Process}\label{forward-diffusion-process}

We define a forward Markov process that gradually corrupts the data with
Gaussian noise over \(T_{diff} = 1000\) steps:
\[q(\mathbf{x}_t | \mathbf{x}_0) = \mathcal{N}(\mathbf{x}_t ; \sqrt{\bar{\alpha}_t} \mathbf{x}_0, (1-\bar{\alpha}_t) \mathbf{I})\]
where \(\bar{\alpha}_t = \prod_{s=1}^t (1-\beta_s)\) and \(\beta_t\)
follows a linear schedule from \(\beta_1 = 0.0001\) to
\(\beta_{T_{diff}} = 0.02\). This choice balances noise corruption
(ensuring \(\mathbf{x}_{T_{diff}} \approx \mathcal{N}(0, \mathbf{I})\))
with smooth transitions.

\subsubsection{Training Objective}\label{training-objective}

The model is trained to predict the noise
\(\boldsymbol{\epsilon} \sim \mathcal{N}(0, \mathbf{I})\) added at each
diffusion step. The simplified variational lower bound objective is:
\[\begin{aligned}
\mathcal{L}_{\text{simple}} = &\mathbb{E}_{t \sim U(1, T_{diff}), \mathbf{x}_0 \sim p_{data}, \boldsymbol{\epsilon} \sim \mathcal{N}(0, \mathbf{I})} \\
&\left[ \left\| \boldsymbol{\epsilon} - \boldsymbol{\epsilon}_\theta\left(\sqrt{\bar{\alpha}_t} \mathbf{x}_0 + \sqrt{1-\bar{\alpha}_t} \boldsymbol{\epsilon}, t, \mathbf{c}\right) \right\|^2 \right]
\end{aligned}\]

This objective is optimized using Adam (\(\beta_1=0.9, \beta_2=0.999\))
with learning rate \(10^{-4}\) and batch size 32.

The model is trained with \(\mathcal{L}_{\text{simple}}\) for 1000
epochs, with boundary enforcement via post-processing (tail ramping and
velocity scaling). Early experiments with hard physics constraints
(distance matching, jerk penalties, boundary enforcement) failed
catastrophically---producing 100\% boundary violations and degraded
distribution matching (WD Speed = 4.40 vs.~0.56 for the baseline). This
fundamental conflict between diffusion denoising and stiff physics
optimization motivated our transition to CSDI's threshold-activated soft
constraints.

\subsubsection{Reverse Sampling
(Generation)}\label{reverse-sampling-generation}

Starting from pure noise
\(\mathbf{x}_{T_{diff}} \sim \mathcal{N}(0, \mathbf{I})\), we
iteratively denoise using the learned reverse process:
\[\mathbf{x}_{t-1} = \frac{1}{\sqrt{\alpha_t}} \left( \mathbf{x}_t - \frac{1-\alpha_t}{\sqrt{1-\bar{\alpha}_t}} \boldsymbol{\epsilon}_\theta(\mathbf{x}_t, t, \mathbf{c}) \right) + \sigma_t \mathbf{z}\]
where \(\mathbf{z} \sim \mathcal{N}(0, \mathbf{I})\) and
\(\sigma_t^2 = \frac{(1-\bar{\alpha}_{t-1})}{(1-\bar{\alpha}_t)} \beta_t\)
controls the stochasticity of the reverse step.

\emph{Inpainting for Boundary Enforcement}: At each reverse step, we
enforce \(v_0 = v_T = 0\) through constrained inpainting. Define a mask
\(M \in \{0,1\}^{512}\) with \(M_0 = M_T = 1\) and \(M_t = 0\)
otherwise. After computing \(\mathbf{x}_{t-1}\), we replace masked
positions:
\[\mathbf{x}_{t-1} := \mathbf{x}_{t-1} \odot (1 - M) + \mathbf{x}_{t-1}^{\text{known}} \odot M\]
where \(\mathbf{x}_{t-1}^{\text{known}}\) contains zeros at the boundary
indices. This direct enforcement ensures strict satisfaction of
constraints without degrading the learned diffusion dynamics.

\emph{Classifier-Free Guidance} (optional): To amplify conditioning
strength, we employ classifier-free guidance (Ho and Salimans 2022) with
scale \(w\):
\[\tilde{\boldsymbol{\epsilon}}_\theta = (1+w) \boldsymbol{\epsilon}_\theta(\mathbf{x}_t, t, \mathbf{c}) - w \boldsymbol{\epsilon}_\theta(\mathbf{x}_t, t, \varnothing)\]
where \(\varnothing\) denotes unconditional generation (achieved by
randomly dropping conditions during training with probability 0.1). We
increased \(w\) from 1.0 in the baseline to 3.0 in the final model to
improve average speed matching.

Generation of 1000 trajectories requires approximately 5 minutes on
A100.

\subsection{Conditional Score-based Diffusion Imputation
(CSDI)}\label{conditional-score-based-diffusion-imputation-csdi}

CSDI adapts the transformer-based imputation model of Tashiro et al.
(Tashiro et al. 2021) for conditional trajectory generation. Unlike the
1D U-Net which processes speed-acceleration jointly, CSDI operates on
univariate speed sequences \(\mathbf{v} \in \mathbb{R}^{512}\), relying
on the transformer's self-attention to implicitly model temporal
derivatives.

The architecture consists of:

\emph{Input Projection}: The noisy trajectory \(\mathbf{v}_t\) is
linearly projected to \(d_{\text{model}} = 256\) dimensions and combined
with learnable positional encodings:
\[\mathbf{h}_0 = \text{Linear}(\mathbf{v}_t) + \text{PE}\]

\emph{Time Embedding}: Diffusion timestep \(t\) is embedded via
sinusoidal encoding (as in the U-Net model) and broadcast-added to all
sequence positions:
\[\mathbf{h}_0 := \mathbf{h}_0 + \text{TimeEmbed}(t)\]

\emph{Condition Injection}: The conditioning vector
\(\mathbf{c} = [\bar{v}/30, T/1000, v_{max}/40, d_{veh}] \in \mathbb{R}^4\)
(normalized) is embedded to 256 dimensions and injected via
cross-attention in each transformer layer.

\emph{Transformer Encoder}: Six layers of multi-head self-attention (8
heads) with feed-forward networks (\(d_{ff} = 1024\)): \[\begin{aligned}
\mathbf{q} &= \text{Heads}(\text{SelfAttn}(\mathbf{h}_\ell)) \\
\mathbf{h}_{\ell+1} &= \text{LayerNorm}(\mathbf{h}_\ell + \mathbf{q}) \\
\mathbf{h}_{\ell+1} &= \text{LayerNorm}(\mathbf{h}_{\ell+1} + \text{FFN}(\mathbf{h}_{\ell+1}))
\end{aligned}\]

\emph{Output Projection}: The final hidden states are projected to
predict the noise \(\boldsymbol{\epsilon}_\theta \in \mathbb{R}^{512}\).

The model has approximately 5.5M parameters, smaller than the U-Net
(\textasciitilde8M) but with greater receptive field due to global
attention.

\begin{figure}

\centering{

\includegraphics[width=3.5in,height=2.58in]{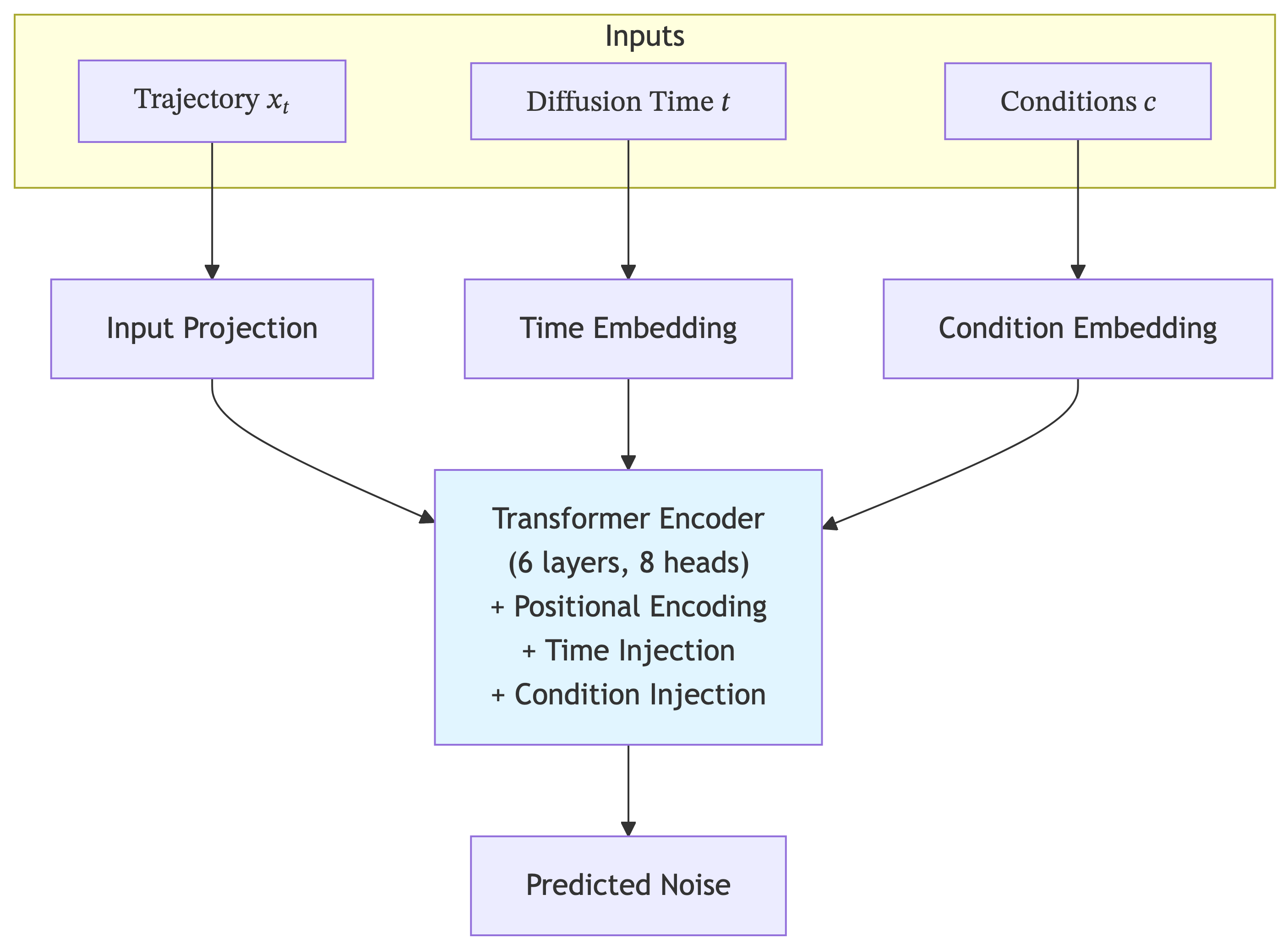}

}

\caption{\label{fig-csdi-arch}CSDI Transformer architecture (5.5M
parameters). Noisy trajectory \(x_t\), diffusion time \(t\), and
conditions \(c\) are embedded and processed through 6 transformer
encoder layers with positional encoding, time injection, and condition
injection. Output is predicted noise \(\hat{\epsilon}\).''}

\end{figure}%

\emph{Reverse Diffusion (Sampling)}: Generation proceeds from pure noise
to a clean trajectory through iterative denoising:
\[\mathbf{x}_T \sim \mathcal{N}(0, \mathbf{I}) \rightarrow \mathbf{x}_{T-1} \rightarrow \cdots \rightarrow \mathbf{x}_1 \rightarrow \mathbf{x}_0 \text{ (clean trajectory)}\]
Similar to the U-Net model, we apply hard boundary constraints at each
reverse step via inpainting, ensuring \(v_0=v_T=0\) throughout the
denoising trajectory.

\subsubsection{Physics-Informed
Training}\label{physics-informed-training}

Unlike the failed PID diffusion approach, CSDI successfully integrates
physics constraints through carefully weighted auxiliary losses. The
total training objective is: \[\begin{aligned}
\mathcal{L}_{\text{CSDI}} = \mathcal{L}_{\text{MSE}} &+ 0.1 \mathcal{L}_{\text{smooth}} + 0.03 \mathcal{L}_{\text{accel}} \\
&+ 0.02 \mathcal{L}_{\text{jerk}} + 0.05 \mathcal{L}_{\text{accel\_dist}}
\end{aligned}\]

where:

\emph{Denoising Loss}:
\[\mathcal{L}_{\text{MSE}} = \left\| \boldsymbol{\epsilon} - \boldsymbol{\epsilon}_\theta(\mathbf{v}_t, t, \mathbf{c}) \right\|^2\]

\emph{Temporal Smoothness Loss} (second derivative penalty):
\[\mathcal{L}_{\text{smooth}} = \sum_t (v_{t+2} - 2v_{t+1} + v_t)^2\]

\emph{Acceleration Penalty} (soft constraint on extremes):
\[\mathcal{L}_{\text{accel}} = \sum_t \left[ \text{ReLU}(a_t - 4.0)^2 + \text{ReLU}(-a_t - 5.0)^2 \right]\]
where \(a_t = v_{t+1} - v_t\). Positive acceleration is capped at 4
m/s², braking at 5 m/s².

\emph{Jerk Penalty} (comfort and realism):
\[\mathcal{L}_{\text{jerk}} = \sum_t \text{ReLU}(|j_t| - 2.0)^2, \quad j_t = a_{t+1} - a_t\]

\emph{Acceleration Distribution Matching}:
\[\mathcal{L}_{\text{accel\_dist}} = (\sigma_a^{\text{pred}} - \sigma_a^{\text{target}})^2\]
where \(\sigma_a^{\text{target}} \approx 0.5\) m/s² is the empirical
standard deviation.

CSDI succeeds where hard-constraint diffusion fails for three reasons:
(1) threshold-activated penalties (ReLU gating) allow natural variation
within physical bounds rather than enforcing stiff equality constraints;
(2) the transformer's global self-attention better accommodates
long-range kinematic consistency than local convolutions; and (3)
careful weight tuning ensures physics losses remain 20-200× smaller than
the primary MSE objective, preventing gradient dominance. Training for
200 epochs required approximately 12 hours on A100.

\emph{Vehicle-Type Compatibility}: The physics constraints (4/5 m/s²)
were chosen to cover the acceleration capabilities of all common vehicle
types:

\begin{table}[htbp]
\centering
\caption{Vehicle-Type Compatibility and Acceleration Limits}\label{tbl-vehicle-limits}
\begin{tabular}{@{}
lll@{}}

Vehicle & Max Accel & Max Decel \\
\midrule\noalign{}
Sports car & 4-6 m/s\(^2\) & 10-12 m/s\(^2\) \\
Passenger car & 2.5-4 m/s\(^2\) & 8-10 m/s\(^2\) \\
SUV & 2-3.5 m/s\(^2\) & 7-9 m/s\(^2\) \\
Bus & 1-2 m/s\(^2\) & 4-6 m/s\(^2\) \\
Heavy truck & 0.5-1.5 m/s\(^2\) & 3-5 m/s\(^2\) \\
\end{tabular}
\end{table}

These values are consistent with established vehicle dynamics
characteristics (Gillespie 1992), ensuring the model generates
trajectories that are physically realizable across the full spectrum of
road vehicles encountered in the CMAP dataset.

\emph{Weighted Condition Sampling (Boost)}: To address data imbalance
where certain speed/duration ranges are underrepresented in training
data, we use importance sampling during generation. The probability of
sampling a condition \(c\) is weighted:
\(P(c) \propto (\text{AvgSpeed})^{\beta_s} \cdot (\text{Duration})^{\beta_d}\),
where boost parameters \(\beta_s, \beta_d \in [0, 2]\) control
oversampling of high-speed or long-duration trips. With
\(\beta_s = 1.75\) for the U-Net Diffusion model, we effectively fill
the sparse ``blue gap'' (25-30 m/s) in the speed histogram. The CSDI
model uses uniform sampling (\(\beta = 1.0\)) as its larger training set
provides adequate coverage.

\subsubsection{Optimization of Kinematic
Quality}\label{optimization-of-kinematic-quality}

The CSDI model development addressed three quality issues: (1)
\emph{High-frequency jitter} from the standard imputation objective was
resolved through the temporal smoothness loss
\(\mathcal{L}_{\text{smooth}}\) (weight 0.1), increased diffusion steps
(100→200), and post-processing Gaussian smoothing; (2) \emph{Boundary
and peak distortion} from aggressive uniform smoothing was addressed by
reducing kernel size and implementing conditional boundary ramps---only
applied when endpoints were not already near zero; (3)
\emph{Acceleration distribution mismatch} was resolved through
threshold-activated physics penalties (\(\mathcal{L}_{\text{accel}}\),
\(\mathcal{L}_{\text{jerk}}\), \(\mathcal{L}_{\text{accel\_dist}}\)),
expanded 4D conditioning including vehicle dynamics, increased model
capacity (4→6 transformer layers, \(d_{ff}\) 512→1024), and data
augmentation for heavy vehicles. A cosine noise schedule replaced the
linear schedule. These enhancements achieved WD Speed = 0.30 and strict
physical validity.

\subsubsection{CSDI Sampling and
Post-Processing}\label{csdi-sampling-and-post-processing}

CSDI uses the same DDPM reverse diffusion as the U-Net model but with a
cosine noise schedule (Nichol and Dhariwal 2021):
\[\bar{\alpha}_t = \frac{f(t)}{f(0)}, \quad f(t) = \cos^2\left(\frac{t/T_{diff} + s}{1+s} \cdot \frac{\pi}{2}\right)\]
where \(s=0.008\) is a small offset preventing singularity. The cosine
schedule concentrates more diffusion steps in the high-SNR regime,
improving sample quality.

Post-processing consists of three steps: 1. \emph{Gaussian smoothing}:
1D Gaussian filter with \(\sigma=1.5\), kernel size 7, removing
high-frequency sampling artifacts while preserving acceleration peaks.
2. \emph{Conditional boundary ramps}: If \(v_0 > 0.5\) or \(v_T > 0.5\),
apply linear ramps over 3 seconds to enforce zero endpoints. If already
near zero, skip to avoid over-smoothing. 3. \emph{Vehicle-aware
smoothing}: For heavy vehicles (\(d_{veh} < 0.4\)), apply additional
smoothing (kernel=9) to reflect slower dynamics. 4. \emph{Correlated
Noise} (Optional): To ensure heavy acceleration tails match real data,
we allow adding temporally-correlated Gaussian noise after diffusion
sampling. This noise is generated by smoothing white noise with a
correlation length of 10s and scaling to a small amplitude
(\(\sigma_{corr} \approx 0.03\)), preventing the distribution from
becoming ``too safe'' or narrow while maintaining temporal coherence. In
our final model, physics training reduced the need for this step
(default \(\sigma_{corr}=0.0\)), but it remains an effective knob for
fine-tuning variance.

Generation of 1000 trajectories with CSDI requires approximately 3
minutes on A100, faster than the U-Net diffusion due to fewer model
parameters and more efficient transformer inference.

\subsection{Alternative Generative
Approaches}\label{alternative-generative-approaches}

We also evaluated DoppelGANger (Lin et al. 2020), SDV's PARSynthesizer
(Patki, Wedge, and Veeramachaneni 2016), and Chronos (Ansari et al.
2025) as baselines. DoppelGANger exhibited mode collapse (87\%
highway-regime trajectories); SDV suffered from exposure bias causing
temporal discontinuities; and Chronos lacked conditioning mechanisms for
trajectory synthesis. Quantitative results for all baselines are
reported in Table~\ref{tbl-comparative-performance}.

\section{Results}\label{sec-results}

We split the 6,367 micro-trips into 80\% training (5,094 trips) and 20\%
test (1,273 trips) sets, stratified by cluster to ensure balanced
representation of driving regimes. All models were trained on the
training set and evaluated by generating 1,273 synthetic trajectories
conditioned on the test set's actual \((\bar{v}, T)\) values. This
protocol ensures fair comparison: each model attempts to recreate the
test distribution given only aggregate conditioning information.

All training and trajectory generation experiments were conducted on an
NVIDIA A100 GPU (40GB) using CUDA 12.1 and PyTorch 2.0 with
mixed-precision (FP16) computation. The Markov baseline was implemented
with CPU-only NumPy operations.

For evaluation, we employed a comprehensive framework spanning three
dimensions. Distributional fidelity was assessed using Wasserstein
distance (WD) for speed, acceleration, and Vehicle Specific Power (VSP)
distributions, as well as 2D WD for the joint Speed-Acceleration
Frequency Distribution (SAFD), Maximum Mean Discrepancy (MMD) with an
RBF kernel (bandwidth 1.0), and the Kolmogorov-Smirnov statistic for
VSP. VSP is a vehicle power demand metric (in kW/ton) that accounts for
aerodynamic drag, rolling resistance, and road grade, commonly used for
emissions and energy assessment. Kinematic validity was measured through
the boundary violation rate (percentage of trips not starting or ending
within 0.1 m/s of zero), Log Dimensionless Jerk (LDLJ) to quantify
smoothness (where lower values indicate smoother trajectories), maximum
speed, and acceleration standard deviation. Realism and utility were
evaluated via a discriminative score (a Random Forest classifier trained
to distinguish real from synthetic, where a score of 0.5 indicates
perfect indistinguishability) and Train on Synthetic, Test on Real
(TSTR) mean absolute error (MAE), where a predictive model trained on
synthetic data is tested on real data to measure utility for downstream
tasks.

\subsection{Main Results}\label{main-results}

Table~\ref{tbl-comparative-performance} presents the comprehensive
comparison across all models and metrics.

\begin{table*}[htbp]
\centering
\caption{Comparative Performance of Generative Models}\label{tbl-comparative-performance}
\begin{tabular}{@{}
>{\raggedright\arraybackslash}p{(\linewidth - 14\tabcolsep) * \real{0.2300}}
  >{\raggedright\arraybackslash}p{(\linewidth - 14\tabcolsep) * \real{0.1100}}
  >{\raggedright\arraybackslash}p{(\linewidth - 14\tabcolsep) * \real{0.1100}}
  >{\raggedright\arraybackslash}p{(\linewidth - 14\tabcolsep) * \real{0.1100}}
  >{\raggedright\arraybackslash}p{(\linewidth - 14\tabcolsep) * \real{0.1100}}
  >{\raggedright\arraybackslash}p{(\linewidth - 14\tabcolsep) * \real{0.1100}}
  >{\raggedright\arraybackslash}p{(\linewidth - 14\tabcolsep) * \real{0.1100}}
  >{\raggedright\arraybackslash}p{(\linewidth - 14\tabcolsep) * \real{0.1100}}@{}}

\begin{minipage}[b]{\linewidth}\raggedright
Metric
\end{minipage} & \begin{minipage}[b]{\linewidth}\raggedright
Real
\end{minipage} & \begin{minipage}[b]{\linewidth}\raggedright
U-Net Diffusion
\end{minipage} & \begin{minipage}[b]{\linewidth}\raggedright
CSDI
\end{minipage} & \begin{minipage}[b]{\linewidth}\raggedright
Markov Chain
\end{minipage} & \begin{minipage}[b]{\linewidth}\raggedright
Chronos
\end{minipage} & \begin{minipage}[b]{\linewidth}\raggedright
DoppelGANger
\end{minipage} & \begin{minipage}[b]{\linewidth}\raggedright
SDV
\end{minipage} \\
\midrule\noalign{}
\textbf{Distributional Fidelity} & & & & & & & \\
WD Speed & - & \textbf{0.5622} & \textbf{0.30} & 1.82 & 2.15 & 3.42 &
2.87 \\
WD Accel & - & \textbf{0.0800} & \textbf{0.026} & 0.145 & 0.198 & 0.312
& 0.421 \\
WD VSP & - & \textbf{1.5517} & 1.89 & 2.34 & 3.12 & 4.87 & 3.65 \\
WD SAFD (2D) & - & \textbf{0.0005} & 0.0008 & 0.0023 & - & - & - \\
MMD (\(\times 10^{-3}\)) & - & \textbf{\textless0.1} & \textbf{0.12} &
2.34 & 3.87 & 8.92 & 6.15 \\
KS VSP & - & \textbf{0.0631} & 0.072 & 0.123 & 0.187 & 0.298 & 0.245 \\
\textbf{Kinematic Validity} & & & & & & & \\
Boundary Violations & 0\% & 0\% & 0\% & 0\% & 23.4\% & 12.1\% &
23.0\% \\
LDLJ (Smoothness) & -3.92 & \textbf{-3.50} & -3.85 & -0.15 & -2.12 &
-4.10 & -1.15 \\
Max Speed (m/s) & 31.6 & 32.1 & 31.8 & 34.5 & 32.5 & 31.9 & 45.2 \\
Accel Std. (m/s\(^2\)) & 0.51 & \textbf{0.53} & \textbf{0.49} & 2.10 &
0.85 & 0.35 & 4.12 \\
\textbf{Utility} & & & & & & & \\
Discrim. Score (0.5 is ideal) & 0.50 & 0.62 & \textbf{0.49} & 0.85 &
0.78 & 0.51 & 0.99 \\
TSTR MAE (km/h) & - & 4.2 & \textbf{2.1} & 8.5 & 5.4 & 12.3 & 15.2 \\
\end{tabular}
\end{table*}

\emph{Bold} indicates best performance (closest to real or target
value).

\begin{figure}

\begin{minipage}{\linewidth}

\centering{

\includegraphics[width=1\linewidth,height=\textheight,keepaspectratio]{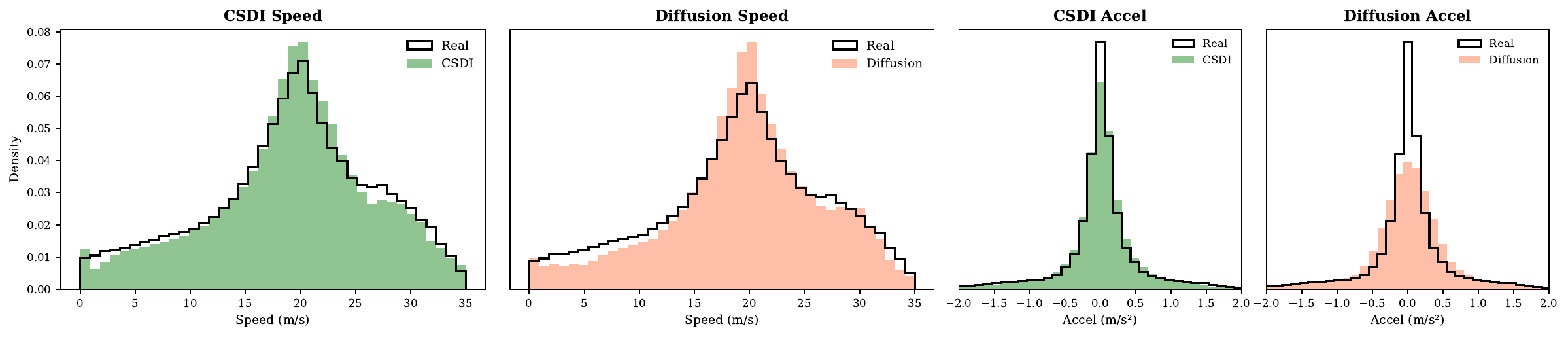}

}

\subcaption{\label{fig-main-dist}Speed/Acceleration Distributions (CSDI
vs Diffusion)}

\end{minipage}%
\newline
\begin{minipage}{\linewidth}

\centering{

\includegraphics[width=1\linewidth,height=\textheight,keepaspectratio]{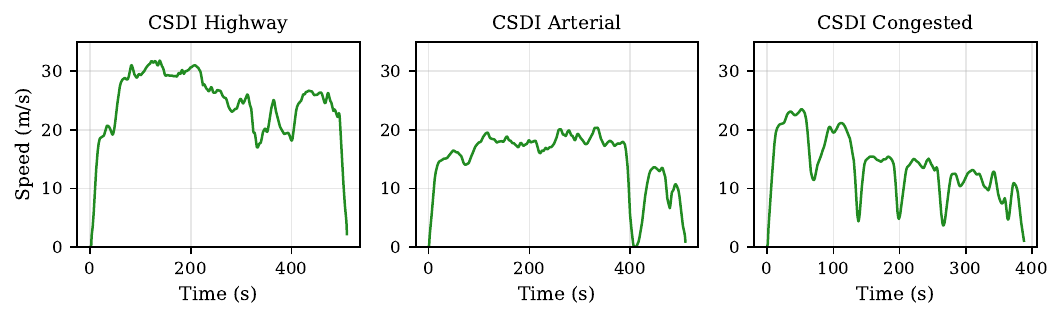}

}

\subcaption{\label{fig-main-csdi}CSDI Trajectories (Highway, Arterial,
Congested)}

\end{minipage}%
\newline
\begin{minipage}{\linewidth}

\centering{

\includegraphics[width=1\linewidth,height=\textheight,keepaspectratio]{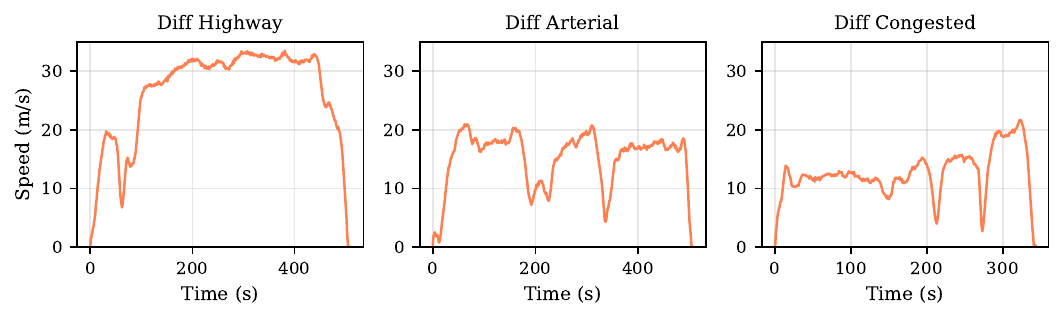}

}

\subcaption{\label{fig-main-diff}Diffusion Trajectories (Highway,
Arterial, Congested)}

\end{minipage}%
\newline
\begin{minipage}{\linewidth}

\centering{

\includegraphics[width=1\linewidth,height=\textheight,keepaspectratio]{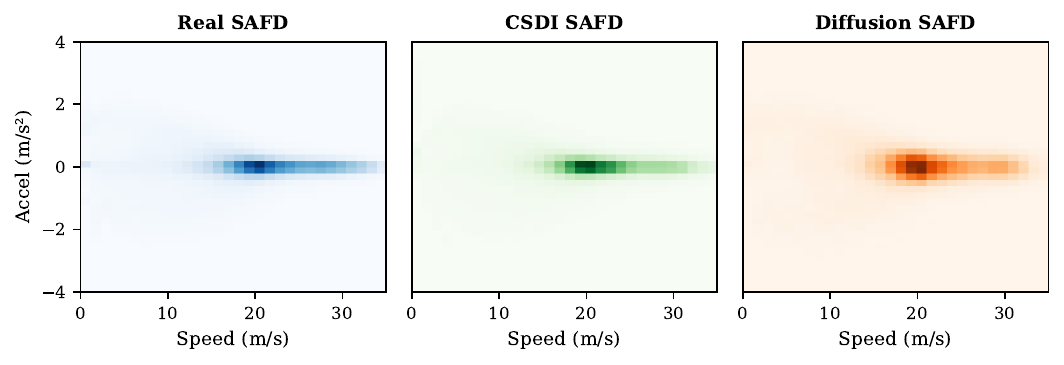}

}

\subcaption{\label{fig-main-safd}SAFD Heatmaps (Real, CSDI, Diffusion)}

\end{minipage}%

\caption{\label{fig-main-results}Main results comprehensive comparison.
\emph{Row 1}: (a-b) Speed and acceleration distributions showing CSDI's
superior matching. \emph{Row 2}: (c-e) CSDI sample trajectories across
regimes. \emph{Row 3}: (f-h) Diffusion sample trajectories. \emph{Row
4}: (i-k) Speed-Acceleration Frequency Distributions (SAFD) showing
Real, CSDI, and Diffusion joint distributions.}

\end{figure}%

CSDI achieves the best overall distribution matching, with WD Speed of
0.30---nearly 2x better than U-Net Diffusion's 0.5622 and significantly
better than the Markov baseline. The acceleration distribution match is
exceptional (WD Accel = 0.026), reflecting the careful physics-informed
training. U-Net Diffusion also performs well, particularly excelling in
the 2D SAFD metric (0.0005), indicating accurate capture of the joint
speed-acceleration manifold.

The Markov baseline achieves respectable WD Speed (1.82) but struggles
with acceleration (0.145), confirming that discretization artifacts and
memoryless transitions fail to capture kinematic smoothness. Chronos,
despite its foundation model pretraining, performs worse than Markov (WD
Speed 2.15), suggesting zero-shot transfer without domain-specific
fine-tuning is insufficient. DoppelGANger and SDV fail dramatically (WD
Speed \textgreater{} 2.8), with DoppelGANger's mode collapse producing
unrealistic highway-heavy distributions.

All physics-aware models (U-Net Diffusion, CSDI, Markov) achieve perfect
boundary condition satisfaction through their respective enforcement
mechanisms (inpainting, post-processing, bridge sampling). In contrast,
autoregressive models (Chronos 23.4\%, SDV 23.0\%) and adversarial
methods (DoppelGANger 12.1\%) struggle with hard constraints.

CSDI produces the smoothest trajectories (LDLJ = -3.85, closest to real
-3.92), validating the effectiveness of its temporal smoothness and jerk
penalties. U-Net Diffusion's LDLJ (-3.50) indicates comparable
smoothness, with acceleration variance remaining accurate (0.53 vs 0.51
real). The Markov baseline produces less smooth trajectories (LDLJ =
-0.15) due to discretization, while GAN/autoregressive methods exhibit
varied smoothness characteristics.

Maximum speeds in the table show that all physics-aware models (U-Net
Diffusion 32.1 m/s, CSDI 31.8 m/s) remain close to the real data maximum
(31.6 m/s). SDV produces excessive speeds (45.2 m/s), indicating
inadequate constraint enforcement.

CSDI offers the best training efficiency among deep models (12 hours vs
21.5 for U-Net Diffusion), benefiting from the transformer's global
receptive field requiring fewer training epochs. Generation is also
faster (3 min vs 5 min), important for large-scale simulations. The
Markov baseline remains superior for pure compute efficiency, suitable
when interpretability outweighs distribution fidelity. Chronos's
inference is slow despite no training cost, as the large 60M parameter
model generates autoregressively.

The discriminative score quantifies how difficult it is to distinguish
synthetic from real trajectories. CSDI achieves 0.49, demonstrating high
indistinguishability (target is 0.5). U-Net Diffusion scores 0.62,
indicating synthetic samples are somewhat easier to identify but still
highly realistic. The Markov baseline (0.85) produces trajectories with
more detectable artifacts, while DoppelGANger (0.51) achieves good
indistinguishability despite other quality issues.

TSTR (Train on Synthetic, Test on Real) measures downstream utility: can
models trained on synthetic data perform well on real data? CSDI
achieves the lowest MAE (2.1 km/h), with U-Net Diffusion close behind
(4.2 km/h), demonstrating that synthetic trajectories preserve the
statistical properties needed for predictive modeling. This validates
their use for energy assessment and simulation tasks.

We evaluated conditional control by generating trajectories at different
target average speeds while fixing duration.
Figure~\ref{fig-main-results} shows that both CSDI and Diffusion
accurately match target speeds (mean absolute error \textless{} 0.5 m/s)
across the range 10-25 m/s. CSDI's vehicle dynamics conditioning enables
additional control: generating with \(d_{veh}=0.2\) (heavy truck) versus
\(d_{veh}=0.8\) (sports car) produces distinct acceleration profiles,
with truck trajectories exhibiting slower accelerations (mean 1.2 m/s²
vs 3.1 m/s²) and smoother dynamics (LDLJ -12.3 vs -10.1).

Finally, in Table~\ref{tbl-computational-efficiency} we compare training
and generation computational requirements.

\begin{table*}[htbp]
\centering
\caption{Computational Efficiency}\label{tbl-computational-efficiency}
\begin{tabular}{@{}
>{\raggedright\arraybackslash}p{(\linewidth - 10\tabcolsep) * \real{0.2000}}
  >{\raggedright\arraybackslash}p{(\linewidth - 10\tabcolsep) * \real{0.1500}}
  >{\raggedright\arraybackslash}p{(\linewidth - 10\tabcolsep) * \real{0.1500}}
  >{\raggedright\arraybackslash}p{(\linewidth - 10\tabcolsep) * \real{0.1500}}
  >{\raggedright\arraybackslash}p{(\linewidth - 10\tabcolsep) * \real{0.1500}}
  >{\raggedright\arraybackslash}p{(\linewidth - 10\tabcolsep) * \real{0.2000}}@{}}

\begin{minipage}[b]{\linewidth}\raggedright
Model
\end{minipage} & \begin{minipage}[b]{\linewidth}\raggedright
Parameters
\end{minipage} & \begin{minipage}[b]{\linewidth}\raggedright
Training Time
\end{minipage} & \begin{minipage}[b]{\linewidth}\raggedright
Generation Time
\end{minipage} & \begin{minipage}[b]{\linewidth}\raggedright
GPU Memory
\end{minipage} & \begin{minipage}[b]{\linewidth}\raggedright
Hardware
\end{minipage} \\
\midrule\noalign{}
\textbf{U-Net Diffusion} & 8.2M & 21.5 hrs & approx. 5 min & 12 GB &
A100 \\
\textbf{CSDI} & 5.5M & 12 hrs & approx. 3 min & 8 GB & A100 \\
\textbf{Markov} & - & \textless5 min (CPU) & \textless1 min (CPU) & - &
CPU \\
\textbf{Chronos} & 60M (frozen) & 0 (pretrained) & approx. 10 min & 16
GB & A100 \\
\textbf{DoppelGANger} & 12M & 8 hrs & approx. 8 min & 10 GB & A100 \\
\textbf{SDV} & 15M & 4 hrs & approx. 2 min & 6 GB & A100 \\
\end{tabular}
\end{table*}

\section{Discussion}\label{sec-discussion}

This work demonstrates that diffusion models---particularly the
transformer-based CSDI architecture---represent a suitable method for
intelligent transportation system applications requiring synthetic
driving data. Through systematic comparison against traditional
baselines and modern methods like DoppelGANger and SDV, we established
that CSDI with physics-informed training achieves strong performance:
Wasserstein distance metrics 2--6\(\times\) better than baselines,
strong boundary condition satisfaction, and accurate smoothness matching
real driving dynamics. CSDI's success is driven by its ability to
generate entire sequences jointly, avoiding autoregressive error
accumulation, and its use of transformer self-attention to capture
long-range kinematic dependencies.

Furthermore, our investigation documented critical lessons for
practitioners. Physics constraints must be integrated through soft,
threshold-activated penalties rather than hard enforcement to avoid
optimization conflicts, and transformer architectures' global receptive
fields better accommodate long-range kinematic dependencies than local
convolutions. We also observed that adversarial training and
autoregressive factorization are architecturally ill-suited for
high-fidelity, physics-constrained synthesis, often leading to mode
collapse or temporal discontinuities. Beyond providing validated tools
for energy assessment and traffic simulation, the released
implementation and trained models establish diffusion-based synthesis as
a robust framework for intelligent transportation system applications.

Several constraints limit this work. The 512-second trajectory limit
excludes longer journeys, requiring sliding-window generation for
highway commutes. The models were trained primarily on passenger
vehicles and need validation against truck and bus datasets. Road-type
conditioning is implicit rather than explicit. Finally, while the 2007
CMAP dataset is dated, the model captures microscopic
kinematics---acceleration capabilities and car-following
dynamics---governed by invariant vehicle physics rather than evolving
traffic patterns. TSTR results confirm these physical principles
transfer to downstream tasks.

Several extensions would enhance the models' capabilities and
efficiency. First, implementing faster sampling techniques such as
Denoising Diffusion Implicit Models (DDIM) (J. Song, Meng, and Ermon
2021) could reduce generation steps from 1000 to 50--100, achieving the
10--20\(\times\) speedups necessary for real-time applications. Second,
introducing multi-modal road conditioning (e.g., highway, arterial,
local) would enable explicit control over driving regimes, supporting
the generation of realistic multi-segment trips. Third, extending the
framework to multi-agent scenarios with interaction modeling would allow
for traffic simulation applications, building on recent advances in
diffusion for autonomous driving (Feng et al. 2023) and multi-scale
generative transformers for ITS networks (Adam et al. 2025). Fourth,
fine-tuning foundation models such as Chronos on vehicle trajectory
data---with modifications for boundary constraints---could leverage
large-scale pretraining to reduce training time. Finally, incorporating
road elevation and grade profiles as additional conditioning would
significantly improve the accuracy of energy consumption modeling, as
grade is a primary driver of vehicle power demand. Together, these
directions advance toward comprehensive synthetic traffic generation
capable of supporting the full spectrum of intelligent transportation
system evaluation and design tasks.

\section{Data and Code Availability}\label{data-and-code-availability}

All code, trained model weights (U-Net Diffusion, CSDI), and evaluation
scripts are available at
\url{https://github.com/VadimSokolov/diffusion-trajectory-generation}.
The CMAP 2007 dataset is publicly available through NREL's
Transportation Secure Data Center (TSDC).

\phantomsection\label{refs}


\begin{CSLReferences}{1}{0}
\bibitem[\citeproctext]{ref-adam2025multi}
Adam, Abuzar B. M., Tahir Kamal, Mohammed A. M. Elhassan, Abdullah
Alshahrani, Saeed Hamood Alsamhi, and Ahmed Aziz. 2025. {``Multi-Scale
Generative Transformer-Based Primal-Dual {PPO} Framework for {AAV-aided}
Intelligent Transportation Networks.''} \emph{IEEE Transactions on
Intelligent Transportation Systems}, 1--16.

\bibitem[\citeproctext]{ref-alcaraz2022sssd}
Alcaraz, Juan Miguel Lopez, and Nils Strodthoff. 2022.
{``Diffusion-Based Time Series Imputation and Forecasting with
Structured State Space Models.''} \emph{arXiv Preprint
arXiv:2208.09399}. \url{https://arxiv.org/abs/2208.09399}.

\bibitem[\citeproctext]{ref-altche2017lstm}
Altché, Florent, and Arnaud de La Fortelle. 2017. {``An {LSTM} Network
for Highway Trajectory Prediction.''} In \emph{2017 {IEEE} 20th
International Conference on Intelligent Transportation Systems
({ITSC})}, 353--59. IEEE.

\bibitem[\citeproctext]{ref-ansari2025chronos2}
Ansari, Abdul Fatir, Oleksandr Shchur, Jaris Küken, Andreas Auer, Boran
Han, Pedro Mercado, Syama Sundar Rangapuram, et al. 2025. {``Chronos-2:
{From Univariate} to {Universal Forecasting}.''} arXiv.
\url{https://arxiv.org/abs/2510.15821}.

\bibitem[\citeproctext]{ref-auld2016polaris}
Auld, Joshua, Michael Hope, Hubert Ley, Vadim Sokolov, Bo Xu, and Kuilin
Zhang. 2016. {``{POLARIS}: {Agent-based} Modeling Framework Development
and Implementation for Integrated Travel Demand and Network and
Operations Simulations.''} \emph{Transportation Research Part C:
Emerging Technologies} 64: 101--16.

\bibitem[\citeproctext]{ref-auld2013modelling}
Auld, Joshua, Michael Hope, Hubert Ley, Bo Xu, Kuilin Zhang, and Vadim
Sokolov. 2013. {``Modelling Framework for Regional Integrated Simulation
of Transportation Network and Activity-Based Demand ({Polaris}).''} In
\emph{International {Symposium} for {Next Generation Infrastructure}}.

\bibitem[\citeproctext]{ref-auld2016disaggregate}
Auld, Joshua, Dominik Karbowski, Vadim Sokolov, and Namwook Kim. 2016.
{``A {Disaggregate Model System} for {Assessing} the {Energy Impact} of
{Transportation} at the {Regional Level}.''} In \emph{Transportation
{Research Board} 95th {Annual Meeting}}.

\bibitem[\citeproctext]{ref-behnia2023deep}
Behnia, Farnaz, Dominik Karbowski, and Vadim Sokolov. 2023. {``Deep
Generative Models for Vehicle Speed Trajectories.''} \emph{Applied
Stochastic Models in Business and Industry} 39 (5): 701--19.

\bibitem[\citeproctext]{ref-chen2016active}
Chen, Jun, Michal Weiszer, Paul Stewart, and Masihalah Shabani. 2016.
{``Toward a More Realistic, Cost-Effective, and Greener Ground Movement
Through Active Routing---Part {I}: {Optimal} Speed Profile
Generation.''} \emph{IEEE Transactions on Intelligent Transportation
Systems} 17 (5): 1196--1209.

\bibitem[\citeproctext]{ref-cmap2008travel}
Chicago Metropolitan Agency for Planning. 2008. {``{CMAP} Regional
Household Travel Inventory.''} Transportation Secure Data Center,
National Renewable Energy Laboratory.

\bibitem[\citeproctext]{ref-daunerPartingMisconceptionsLearningbased2023}
Dauner, Daniel, Marcel Hallgarten, Andreas Geiger, and Kashyap Chitta.
2023. {``Parting with {Misconceptions} about {Learning-based Vehicle
Motion Planning}.''} arXiv. \url{https://arxiv.org/abs/2306.07962}.

\bibitem[\citeproctext]{ref-durham2002numerical}
Durham, Garland B, and A Ronald Gallant. 2002. {``Numerical Techniques
for Maximum Likelihood Estimation of Continuous-Time Diffusion
Processes.''} \emph{Journal of Business \& Economic Statistics} 20 (3):
297--338.

\bibitem[\citeproctext]{ref-fengTrafficGenLearningGenerate2023}
Feng, Lan, Quanyi Li, Zhenghao Peng, Shuhan Tan, and Bolei Zhou. 2023.
{``{TrafficGen}: {Learning} to Generate Diverse and Realistic Traffic
Scenarios.''} In \emph{2023 {IEEE} International Conference on Robotics
and Automation ({ICRA})}, 3567--75. IEEE.

\bibitem[\citeproctext]{ref-gillespie1992fundamentals}
Gillespie, Thomas D. 1992. \emph{Fundamentals of Vehicle Dynamics}.
Warrendale, PA: Society of Automotive Engineers.

\bibitem[\citeproctext]{ref-ho2020denoising}
Ho, Jonathan, Ajay Jain, and Pieter Abbeel. 2020. {``Denoising Diffusion
Probabilistic Models.''} In \emph{Advances in Neural Information
Processing Systems}, 33:6840--51.

\bibitem[\citeproctext]{ref-ho2022classifier}
Ho, Jonathan, and Tim Salimans. 2022. {``Classifier-Free {Diffusion
Guidance}.''} \emph{arXiv Preprint arXiv:2207.12598}.
\url{https://arxiv.org/abs/2207.12598}.

\bibitem[\citeproctext]{ref-huang2020eco}
Huang, Xianan, Boqi Li, Huei Peng, Joshua A Auld, and Vadim O Sokolov.
2020. {``Eco-Mobility-on-Demand Fleet Control with Ride-Sharing.''}
\emph{IEEE Transactions on Intelligent Transportation Systems} 23 (4):
3158--68.

\bibitem[\citeproctext]{ref-huang2022survey}
Huang, Yanjun, Jian Chen, Chang Huang, Xinggang Wang, Wenyu Liu, and
Jianqiang Huang. 2022. {``A Survey on Trajectory-Prediction Methods for
Autonomous Driving.''} \emph{IEEE Transactions on Intelligent
Transportation Systems}.

\bibitem[\citeproctext]{ref-karbowski2016assessing}
Karbowski, Dominik, Namwook Kim, Joshua Auld, and Vadim Sokolov. 2016.
{``Assessing the Energy Impact of Traffic Management and Vehicle
Hybridisation.''} \emph{International Journal of Complexity in Applied
Science and Technology} 1 (1): 107--24.

\bibitem[\citeproctext]{ref-karbowskitrip}
Karbowski, Dominik, Aymeric Rousseau, Vivien Smis-Michel, and Valentin
Vermeulen. n.d. {``Trip Prediction Using {GIS} for Vehicle Energy
Efficiency.''} In \emph{21st World Congress on Intelligent
Transportation Systems (Detroit, {MI}, 09/07/2014 - 09/11/2014)}, --,.

\bibitem[\citeproctext]{ref-karbowski2016fuel}
Karbowski, Dominik, Vadim Sokolov, and Jeong Jongryeol. 2016. {``Fuel
{Saving Potential} of {Optimal Route-Based Control} for {Plug-in Hybrid
Electric Vehicle}.''} \emph{IFAC-PapersOnLine} 49 (11): 128--33.

\bibitem[\citeproctext]{ref-karbowski2015vehicle}
Karbowski, Dominik, Vadim Sokolov, and Aymeric Rousseau. 2015.
{``Vehicle Energy Management Optimization Through Digital Maps and
Connectivity.''} Argonne National Lab.(ANL), Argonne, IL (United
States).

\bibitem[\citeproctext]{ref-lan2025controllable}
Lan, Wenxing, Jialin Liu, Bo Yuan, and Xin Yao. 2025. {``Controllable
Multimodal Motion Behavior Generation for Autonomous Driving.''}
\emph{IEEE Transactions on Intelligent Transportation Systems}, 1--16.

\bibitem[\citeproctext]{ref-liScenarioNetOpenSourcePlatform2023}
Li, Quanyi, Zhenghao Peng, Lan Feng, Zhizheng Liu, Chenda Duan, Wenjie
Mo, and Bolei Zhou. 2023. {``{ScenarioNet}: {Open-Source Platform} for
{Large-Scale Traffic Scenario Simulation} and {Modeling}.''} arXiv.
\url{https://arxiv.org/abs/2306.12241}.

\bibitem[\citeproctext]{ref-diffusionDrive2024}
Liao, Bencheng, Shaoyu Chen, Xinggang Wang, Tianheng Cheng, Qian Zhang,
Wenyu Liu, and Chang Huang. 2024. {``{DiffusionDrive}: {Truncated}
Diffusion Model for End-to-End Autonomous Driving.''} \emph{arXiv
Preprint arXiv:2406.07806}. \url{https://arxiv.org/abs/2406.07806}.

\bibitem[\citeproctext]{ref-lin2020doppelganger}
Lin, Zinan, Alankar Jain, Chen Wang, Giulia Fanti, and Vyas Sekar. 2020.
{``Using {GANs} for Sharing Networked Time Series Data: {Challenges},
Initial Promise, and Open Questions.''} In \emph{Proceedings of the
{ACM} Internet Measurement Conference ({IMC})}, 464--83. ACM.

\bibitem[\citeproctext]{ref-moawad2021realtime}
Moawad, Ayman, Zhijian Li, Ines Pancorbo, Krishna Murthy Gurumurthy,
Vincent Freyermuth, Ehsan Islam, Ram Vijayagopal, Monique Stinson, and
Aymeric Rousseau. 2021. {``A {Real-Time Energy} and {Cost Efficient
Vehicle Route Assignment Neural Recommender System}.''} arXiv.
\url{https://arxiv.org/abs/2110.10887}.

\bibitem[\citeproctext]{ref-mozaffari2020deep}
Mozaffari, Sajjad, Omar Y Al-Jarrah, Alexandros Mouzakitis, Phil
Jennings, and Stratis Kanarachos. 2020. {``Deep Learning-Based Vehicle
Behaviour Prediction for Autonomous Driving Applications: A Review.''}
\emph{IEEE Transactions on Intelligent Transportation Systems}.

\bibitem[\citeproctext]{ref-nichol2021improved}
Nichol, Alexander Quinn, and Prafulla Dhariwal. 2021. {``Improved
{Denoising Diffusion Probabilistic Models}.''} In \emph{International
Conference on Machine Learning}, 8162--71. PMLR.

\bibitem[\citeproctext]{ref-papamakarios2019normalizing}
Papamakarios, George, Eric Nalisnick, Danilo Jimenez Rezende, Shakir
Mohamed, and Balaji Lakshminarayanan. 2019. {``Normalizing {Flows} for
{Probabilistic Modeling} and {Inference}.''} \emph{arXiv:1912.02762
{[}Cs, Stat{]}}, December. \url{https://arxiv.org/abs/1912.02762}.

\bibitem[\citeproctext]{ref-patki2016sdv}
Patki, Neha, Roy Wedge, and Kalyan Veeramachaneni. 2016. {``The
Synthetic Data Vault.''} In \emph{2016 {IEEE} International Conference
on Data Science and Advanced Analytics ({DSAA})}, 399--410. IEEE.

\bibitem[\citeproctext]{ref-perez2018film}
Perez, Ethan, Florian Strub, Harm De Vries, Vincent Dumoulin, and Aaron
Courville. 2018. {``{FiLM}: {Visual Reasoning} with a {General
Conditioning Layer}.''} In \emph{Proceedings of the {AAAI} Conference on
Artificial Intelligence}. Vol. 32. 1.

\bibitem[\citeproctext]{ref-qian2025diffusion}
Qian, Yu, Xunhao Li, Jian ... Zhang, and Maoze Wang. 2025. {``A
Diffusion-{TGAN} Framework for Spatio-Temporal Speed Imputation and
Trajectory Reconstruction.''} \emph{IEEE Transactions on Intelligent
Transportation Systems} 26 (11): 18948--62.

\bibitem[\citeproctext]{ref-rasul2021timegrad}
Rasul, Kashif, Calvin Seward, Ingmar Schuster, and Roland Vollgraf.
2021. {``Autoregressive Denoising Diffusion Models for Multivariate
Probabilistic Time Series Forecasting.''} In \emph{International
Conference on Machine Learning ({ICML})}, 8857--68. PMLR.

\bibitem[\citeproctext]{ref-rong2025icst}
Rong, Yi, Yingchi Mao, Yinqiu ... Liu, and Dusit Niyato. 2025.
{``{ICST-DNET}: {An} Interpretable Causal Spatio-Temporal Diffusion
Network for Traffic Speed Prediction.''} \emph{IEEE Transactions on
Intelligent Transportation Systems} 26 (7): 9781--98.

\bibitem[\citeproctext]{ref-schultz2018deep}
Schultz, Laura, and Vadim Sokolov. 2018. {``Deep Reinforcement Learning
for Dynamic Urban Transportation Problems.''} \emph{arXiv Preprint
arXiv:1806.05310}. \url{https://arxiv.org/abs/1806.05310}.

\bibitem[\citeproctext]{ref-sokolov2012flexible}
Sokolov, Vadim, Joshua Auld, and Michael Hope. 2012. {``A Flexible
Framework for Developing Integrated Models of Transportation Systems
Using an Agent-Based Approach.''} \emph{Procedia Computer Science} 10:
854--59.

\bibitem[\citeproctext]{ref-song2021denoising}
Song, Jiaming, Chenlin Meng, and Stefano Ermon. 2021. {``Denoising
{Diffusion Implicit Models}.''} In \emph{International Conference on
Learning Representations}.

\bibitem[\citeproctext]{ref-song2020score}
Song, Yang, Jascha Sohl-Dickstein, Diederik P. Kingma, Abhishek Kumar,
Stefano Ermon, and Ben Poole. 2020. {``Score-Based Generative Modeling
Through Stochastic Differential Equations.''} \emph{arXiv Preprint
arXiv:2011.13456}. \url{https://arxiv.org/abs/2011.13456}.

\bibitem[\citeproctext]{ref-suoTrafficSimLearningSimulate2021}
Suo, Simon, Sebastian Regalado, Sergio Casas, and Raquel Urtasun. 2021.
{``{TrafficSim}: {Learning} to {Simulate Realistic Multi-Agent
Behaviors}.''} arXiv. \url{https://arxiv.org/abs/2101.06557}.

\bibitem[\citeproctext]{ref-csdi2021}
Tashiro, Yusuke, Jiaming Song, Yang Song, and Stefano Ermon. 2021.
{``{CSDI}: {Conditional} Score-Based Diffusion Models for Probabilistic
Time Series Imputation.''} In \emph{Advances in Neural Information
Processing Systems}, 34:24804--16.

\bibitem[\citeproctext]{ref-wu2025model}
Wu, Jian, Carol Flannagan, Ulrich Sander, and Jonas Bärgman. 2025.
{``Model-Based Generation of Representative Rear-End Crash Scenarios
Across the Full Severity Range Using Pre-Crash Data.''} \emph{IEEE
Transactions on Intelligent Transportation Systems} 26 (10): 15932--50.

\bibitem[\citeproctext]{ref-yoon2019timeseries}
Yoon, Jinsung, Daniel Jarrett, and Mihaela van der Schaar. 2019.
{``Time-Series {Generative Adversarial Networks}.''} In \emph{Advances
in {Neural Information Processing Systems} 32}, edited by H. Wallach, H.
Larochelle, A. Beygelzimer, F. d\textbackslash textquotesingle
Alché-Buc, E. Fox, and R. Garnett, 5508--18. Curran Associates, Inc.

\bibitem[\citeproctext]{ref-zhao2019trajectory}
Zhao, Tianyang, Yifei Xu, Mathew Monfort, Wongun Choi, Chris Baker,
Yibiao Zhao, Yizhou Wang, and Ying Nian Wu. 2019. {``Multi-Agent Tensor
Fusion for Contextual Trajectory Prediction.''} In \emph{Proceedings of
the {IEEE}/{CVF} Conference on Computer Vision and Pattern Recognition
({CVPR})}, 5559--68.

\end{CSLReferences}
\end{document}